\documentclass[sigconf,10pt]{acmart}
\usepackage{graphicx}
\usepackage{subfig}
\usepackage{amsmath}
\usepackage{nicefrac}
\usepackage{listings}
\usepackage{appendix}
\usepackage{xspace}
\usepackage{booktabs} 
\usepackage{balance}  
\usepackage[utf8]{inputenc}
\usepackage{xcolor}
\usepackage{verbatim}
\usepackage[subtle,title=tight]{savetrees}
\PassOptionsToPackage{hyphens}{url}
\usepackage{hyperref}
\usepackage{multirow}
\usepackage{colortbl}
\usepackage{microtype}
\usepackage{algorithm,algorithmicx,algpseudocode}

\renewcommand\footnotetextcopyrightpermission[1]{}
\AtBeginDocument{%
  \providecommand\BibTeX{{%
    \normalfont B\kern-0.5em{\scshape i\kern-0.25em b}\kern-0.8em\TeX}}}

\setcopyright{acmcopyright}
\copyrightyear{2020}
\acmYear{2020}
\acmDOI{xx.xxxx/yyyyyyy.yyyyyyy}

\newcommand{\TheSystem}{Flood\xspace}
\newcommand{\NewPara}[1]{\vspace{4pt}\noindent{\bf #1}}
\newcommand{\Section}[1]{\S\ref{sec:#1}}
\newcommand{\Equation}[1]{Eq.~\ref{eq:#1}}
\newcommand{\Figure}[1]{Fig.~\ref{fig:#1}}
\newcommand{\Table}[1]{Tab.~\ref{tab:#1}}
\newcommand{\Appendix}[1]{Appendix~\ref{app:#1}}

\def\compactify{\itemsep=0pt \topsep=0pt \partopsep=0pt \parsep=0pt \leftmargin=0.5cm}
\let\latexusecounter=\usecounter

\newenvironment{CompactEnumerate} 
  {\def\usecounter{\compactify\latexusecounter} 
   \begin{enumerate}}
  {\end{enumerate}\let\usecounter=\latexusecounter}

\begin{document}


\title{Learning Multi-dimensional Indexes}

\author{Vikram Nathan$^*$, Jialin Ding$^*$, Mohammad Alizadeh, Tim Kraska}
\email{{vikramn,jialind,alizadeh,kraska}@mit.edu}
\affiliation{%
  \institution{Massachusetts Institute of Technology}
}
\authornote{Equal contribution.}


\begin{abstract}

Scanning and filtering over multi-dimensional tables are key operations in modern analytical database engines. To optimize the performance of these operations, databases often create clustered indexes over a single dimension or multi-dimensional indexes such as R-Trees, or use complex sort orders (e.g., Z-ordering). However, these schemes are often hard to tune and their performance is inconsistent across different datasets and queries. In this paper, we introduce \TheSystem, a multi-dimensional in-memory read-optimized index that automatically adapts itself to a particular dataset and workload by jointly optimizing the index structure and data storage layout. \TheSystem achieves up to three orders of magnitude faster performance for range scans with predicates than state-of-the-art multi-dimensional indexes or sort orders on real-world datasets and workloads. Our work serves as a building block towards an end-to-end learned database system.

\end{abstract}

\maketitle

\section{Introduction}
\label{sec:intro}
Scanning and filtering are the foundation of any analytical database engine, and several advances over the past several years specifically target database scan and filter performance.
Most importantly, column stores \cite{cstore} have been proposed to delay or entirely avoid accessing columns (i.e., attributes) which are not relevant to a query. 
Similarly, there exist many techniques to skip over records that do not match a query filter. 
For example, transactional database systems create a clustered B-Tree index on a single attribute, while column stores often sort the data by a single attribute.
The idea behind both is the same: if the data is organized according to an attribute that is present in the query filter, the execution engine can either traverse the B-Tree or use binary search, respectively, to quickly narrow its search to the relevant range in that attribute.
We refer to both approaches as clustered column indexes.

If data has to be filtered by more than one attribute, secondary indexes can be used. 
Unfortunately, their large storage overhead and the latency incurred by chasing pointers make them viable only for a rather narrow use case, namely when the predicate on the indexed attribute has a very high selectivity; in most other cases, scanning the entire table can be faster and more space efficient \cite{vertica}.
An alternative approach is to use \emph{multi-dimensional} indexes to organize the data; these may be tree-based data structures (e.g., k-d trees, R-Trees, or octrees)
or a specialized sort order over multiple attributes (e.g., a space-filling curve like Z-ordering or hand-picked hierarchical sort). 
Indeed, many state-of-the-art analytical database systems use multi-dimensional indexes or sort-orders to improve the scan performance of queries with predicates over several columns.
For example, both Redshift~\cite{redshift} and SparkSQL~\cite{spark-sql} use Z-ordering to lay out the data; Vertica can define a sort-order over multiple columns (e.g., first age, then date), while IBM Informix, along with other spatial database systems, uses an R-Tree~\cite{ibm-rtree}.

However, multidimensional indexes still have significant drawbacks. 
First, these techniques are extremely hard to tune.
For example, Vertica's ability to sort hierarchically on multiple attributes requires an admin to carefully pick the sort order. The admin must therefore know which columns are accessed together, and their selectivity, to make an informed decision.
~
Second, there is no single approach (even if tuned correctly) that dominates all others. 
As our experiments will show, the best multidimensional index varies depending on the data distribution and query workload.
~
Third, most existing techniques cannot be fully tailored for a specific data distribution and query workload. 
While all of them provide tunable parameters (e.g., page size), they do not allow finer-grained customization for a specific dataset and filter access pattern. 

To address these shortcomings, we propose \TheSystem, the first learned multi-dimensional in-memory index.
\TheSystem's goal is to locate records matching a query filter faster than existing indexes, by automatically co-optimizing the data layout and index structure for a particular data and query distribution.

Central to \TheSystem are two key ideas. First, \TheSystem uses a sample query filter workload to learn how often certain dimensions are used, which ones are used together, and which are more selective than others.
Based on this information, \TheSystem automatically customizes the entire layout to optimize query performance on the given workload.
Second, \TheSystem uses empirical CDF models to project the multi-dimensional and potentially skewed data distribution into a more uniform space. 
This ``flattening'' step helps limit the number of points that are searched and is key to achieving good performance.

\TheSystem's learning-based approach to layout optimization is what distinguishes it from other multi-dimensional index structures. It allows \TheSystem to target its performance to a particular query workload, avoid the superlinear growth in index size that plagues some indexes even with uniformly distributed data~\cite{zvalue}, and locate relevant records quickly without the high traversal times incurred by k-d trees and hyperoctrees, especially for larger range scans.

While \TheSystem's techniques are general and may potentially benefit a wide range of systems, from OLTP in-memory transaction processing systems to disk-based data warehouses, this paper focuses on improving multi-dimensional index performance (i.e., reducing unnecessary scan and filter overhead) for an in-memory column store. 
In-memory stores are increasingly popular due to lower RAM prices~\cite{cheap-ram} and the increasing amount of main memory which can be put into a single machine~\cite{exasol, fujitsu}.
In addition, \TheSystem is optimized for reads (i.e., query speed) at the expense of writes (i.e., incremental index updates), making it most suitable for rather static analytical workloads, though our experiments show that adjusting to a new query workload is relatively fast.
We envision that \TheSystem could serve as the building block for a multi-dimensional in-memory key-value store or be integrated into commercial in-memory (offline) analytics accelerators like Oracle's Database In-Memory (DBIM)~\cite{dbim}.

The ability to self-optimize allows \TheSystem to outperform alternative state-of-the-art techniques by up to three orders of magnitude, while often having a significantly smaller storage overhead.
More importantly though, \TheSystem achieves \emph{optimality across the board}: it has better, or at least on-par, performance compared to the next-fastest indexing technique on all our datasets and workloads. 
For example, on a real sales dataset, \TheSystem achieves a boost of $3\times$ over a tuned clustered column index  and $72\times$ over Amazon Redshift's Z-encoding method. On a different workload derived from TPC-H, \TheSystem is $61\times$ faster than the clustered column index but only $3\times$ faster than the Z-encoding.

We make the following contributions:
\begin{CompactEnumerate}
\item We design and implement \TheSystem, the first learned multi-dimensional index, on an in-memory column store. \TheSystem targets its layout for a particular workload by learning from a sample filter predicate distribution.
\item We evaluate a wide range of multi-dimensional indexes on one synthetic and three real-world datasets, including one with a workload from an actual sales database at a major analytical database company. Our evaluation shows that \TheSystem outperforms all other index structures. 
\item We show that \TheSystem achieves query speedups on different filter predicates and data sizes, and its index creation time is competitive with existing multi-dimensional indexes.
\end{CompactEnumerate}

\section{Related Work}
\label{sec:related}

There is a rich corpus of work dedicated to multi-dimensional indexes, and many commercial database systems have turned to multi-dimensional indexing schemes. 
For example, Amazon Redshift organizes points by Z-order~\cite{z-order}, which maps multi-dimensional points onto a single dimension for sorting~\cite{redshift, oracle-zorder, amazon-zorder}. With spatial dimensions, SQL Server allows Z-ordering~\cite{sql-server}, and IBM Informix uses an R-Tree~\cite{ibm-rtree}. 
Other multi-dimensional indexes include  K-d trees, octrees, R$^*$ trees,  UB trees (which also make use of the Z-order), among many others (see \cite{Ooi_indexingspatial,spatialindexsurvey} for a survey).
\TheSystem's underlying index structure is perhaps most similar to Grid Files~\cite{gridfile}, which has many variants~\cite{excell, twolevel-gridfile, twin-gridfile}.
However, Grid Files do not automatically adjust to the query workload, yielding poorer performance (\Section{eval}). In fact, Grid Files tend to have superlinear growth in index size even for uniformly distributed data~\cite{zvalue}.

\TheSystem also differs from other adaptive indexing techniques such as database cracking~\cite{db-cracking, partial-cracking, Schuhknecht:2013:UPD:2732228.2732229}. 
The main goal of cracking is to build a query-adaptive incremental index by partitioning the data incrementally with each observed query.
However, cracking produces only single dimensional clustered indexes, and does not jointly optimize the layout over multiple attributes. This limits its usefulness on queries with multi-dimensional filters. Furthermore, cracking does not take the data distribution into account and adapts only to queries; on the other hand, \TheSystem adapts to both the queries \emph{and} the underlying data.

Arguably most relevant to this work is automatic index selection~\cite{self-driving, self-driving2, self-driving3}. 
However, these approaches mainly focus on creating secondary indexes, whereas \TheSystem optimizes the storage and index itself for a given workload and data distribution.

For aggregation queries, data cubes~\cite{datacube} are an alternative to indexes. 
However, data cubes alone are insufficient for queries over arbitrary filter ranges, and they cannot support arbitrary actions over the queried records (e.g., returning the records themselves).

Finally, learned models have been used to replace/enhance traditional B-trees~\cite{kraska,atree,alex} and secondary indexes~\cite{correlationindex,correlationmaps}. Self-designing systems use learned cost models to synthesize the optimal algorithms for a data structure, resulting in a continuum of possible designs that form a ``periodic table'' of data structures~\cite{data-calculator}.
\TheSystem extends these works in two ways.
First, \TheSystem learns models for indexing \emph{multiple} dimensions. Since there is no natural sort order for points in many dimensions, \TheSystem requires a design tailored specifically to multi-dimensional data. Second, prior work focused solely on constructing models of the data, without taking queries into account. \TheSystem optimizes its layout by learning from the query workload as well. Also unlike~\cite{data-calculator}, \TheSystem embeds models into the data structure itself.

SageDB~\cite{sagedb} proposed the idea of a learned multi-dimensional index but did not describe any details.
\begin{figure}[t!]
\centering
\includegraphics[width=0.8\columnwidth]{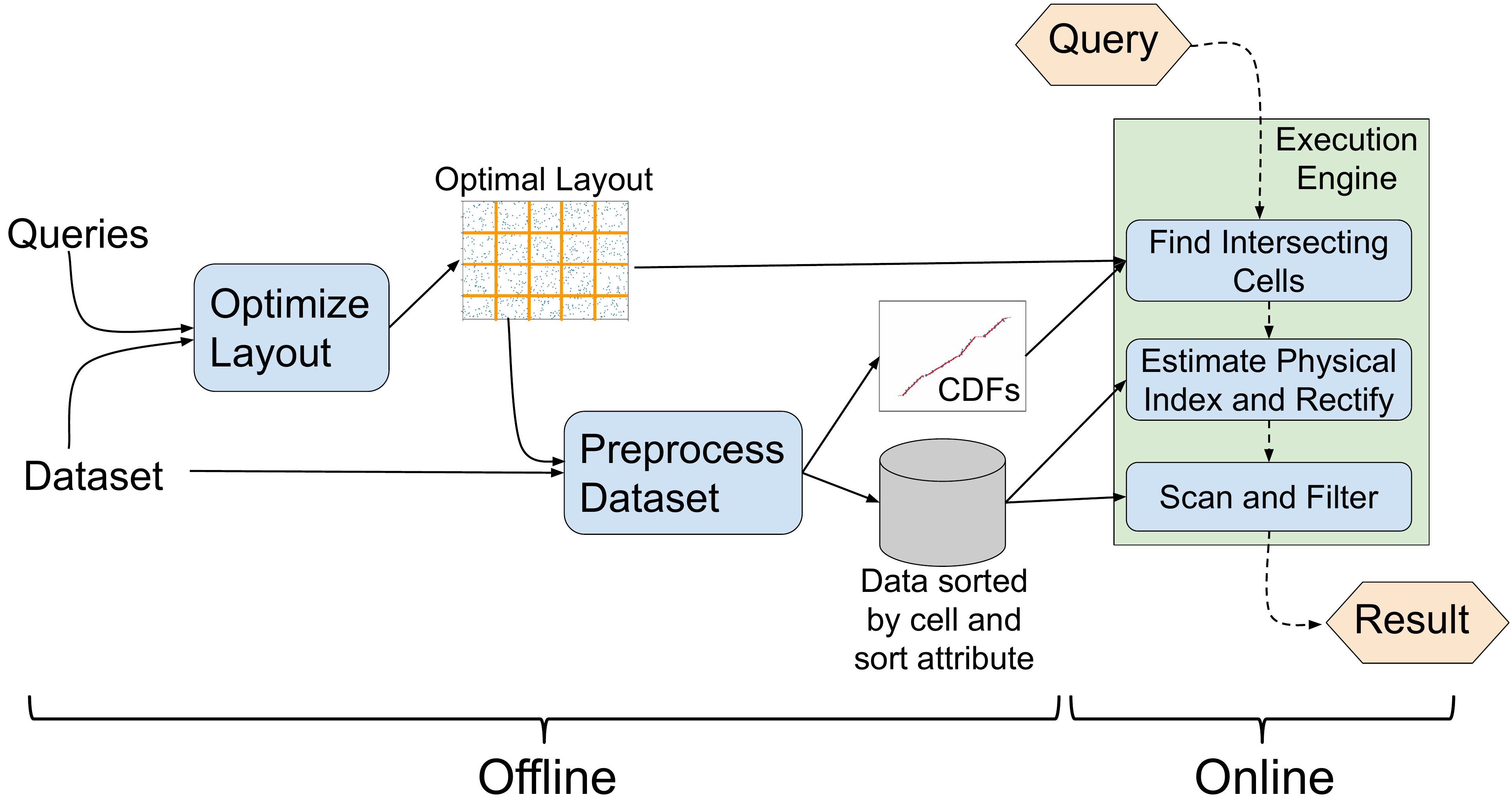}
\vspace{-1em}
\caption{\TheSystem's system architecture.}
\vspace{-2em}
\label{fig:index-arch}
\end{figure}

\section{Index Overview}
\label{sec:design}

\TheSystem is a multi-dimensional clustered index that speeds up the processing of relational queries that select a range over one or more attributes. For example:

\indent\indent\texttt{{\small SELECT SUM(R.X) \\ \indent\indent FROM MyTable \\ \indent\indent WHERE (a $\leq$ R.Y $\leq$ b) AND (c $\leq$ R.Z $\leq$ d)}}

Note that equality predicates of the form \texttt{{\small R.Z == f}} can be rewritten as \texttt{{\small f $\leq$ R.Z $\leq$ f}}.
Typical selections generally also include disjunctions (i.e. \texttt{{\small OR}} clauses). However, these can be decomposed into multiple queries over disjoint attribute ranges; hence our focus on \texttt{{\small AND}}s. 

\TheSystem consists of two parts: (1) an offline preprocessing step that chooses
an optimal layout, creating an index based on that layout, and (2) an online component responsible for executing queries
as they arrive (see \Figure{index-arch}).

At a high level, \TheSystem is a variant of a basic grid index that divides $d$-dimensional data space into a $d$-dimensional grid of contiguous cells, so that data in each cell is stored together. We describe \TheSystem's grid layout and online operation in \Section{design:layout} and \Section{design:queryflow}.
We then discuss \TheSystem's central idea: how to automatically optimize the grid layout's parameters for a particular query workload (\Section{opt}).
The rest of this paper uses the terms \emph{attribute} and \emph{dimension} interchangeably, as well as the terms \emph{record} and \emph{point}.

\subsection{Data Layout}
\label{sec:design:layout}
Consider an index on $d$ dimensions. Unlike the single dimensional case, points in multiple dimensions have no natural sort order. Our first goal is then to impose an ordering over the data.

We first rank the $d$ attributes. The details of how to choose a ranking are discussed in \Section{opt}, but for the purposes of illustration, we assume it is given.
Next, we use the first $d-1$ dimensions in the ordering to overlay a \mbox{($d-1$)-dimensional} grid on the data, where the $i$th dimension in the ordering is divided into $c_i$ equally spaced columns between its minimum and maximum values.
Every point maps to a particular \emph{cell} in this grid, i.e. a tuple with $d-1$ attributes. In particular, if $M_i$ and $m_i$ are the maximum and minimum values of the data along the $i$th dimension, then define the dimension's range as $r_i = M_i - m_i + 1$. Then the cell for point $p = (p_1, \ldots, p_d)$ is:
\[ \text{cell}(p) = \left(\left\lfloor\frac{p_1 - m_1}{r_1}\cdot c_1\right\rfloor, \ldots, \left\lfloor \frac{p_{d-1} - m_{d-1}}{r_{d-1}} \cdot c_{d-1} \right\rfloor\right)\]
Note that the cell is determined only by the first $d-1$ dimensions; the $d$th dimension, the \emph{sort dimension}, will be used to order points within a cell.

\begin{figure}[t!]
\centering
\includegraphics[width=0.7\columnwidth,height=10em]{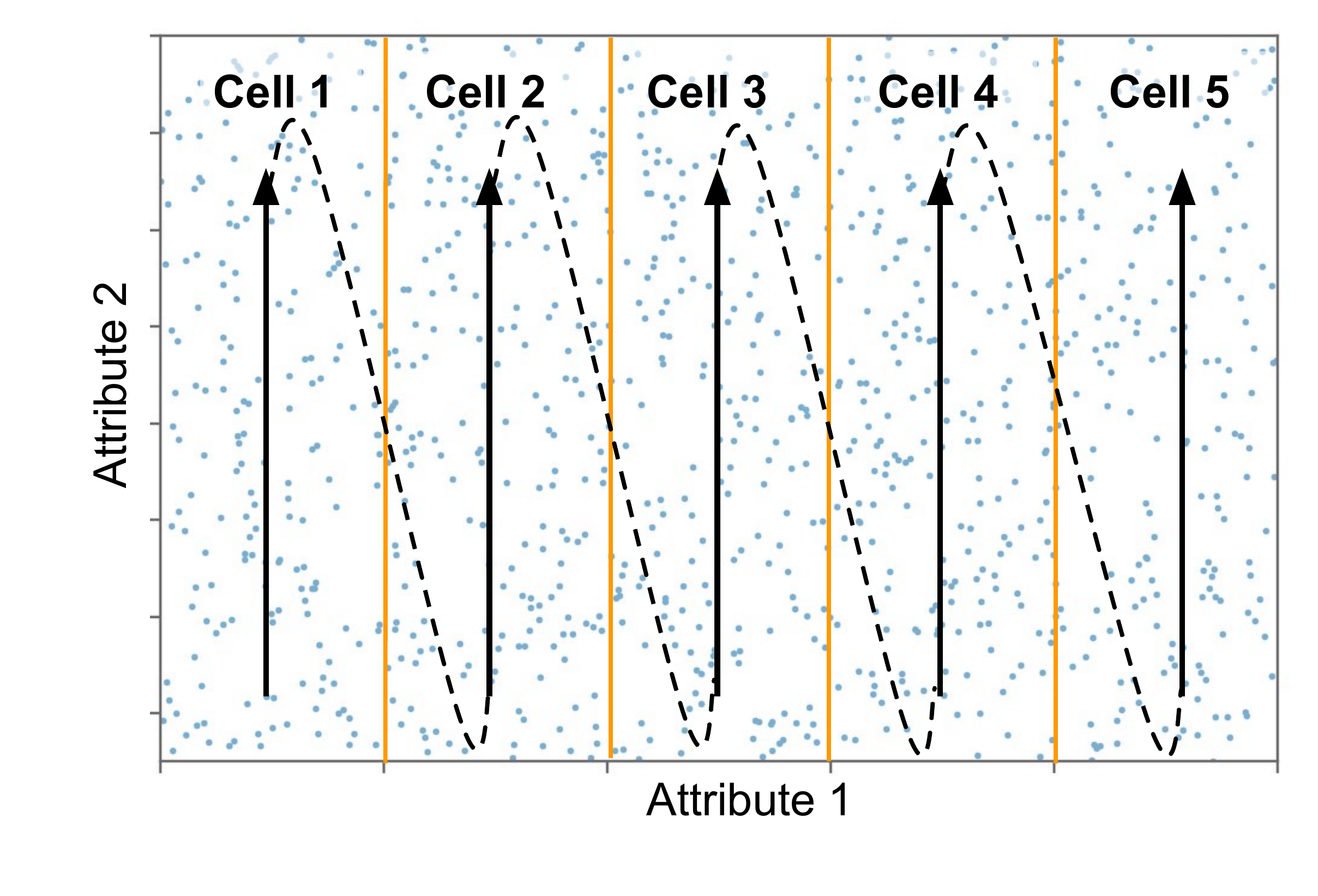}
\caption{A basic layout in 2D, with dimension order (x, y) and $c_0 = 5$. Points are bucketed into columns along x and then sorted by their y-values, creating the seriliaziation order indicated by the arrows.}
\vspace*{-10pt}
\label{fig:2d-naive-layout}
\end{figure}

\TheSystem orders the points using a depth-first traversal of the cells along the dimension ordering, i.e. cells are sorted by the first value in the tuple, then the second, etc.  Within each cell, points are sorted by their value in the $d$th dimension. \Figure{2d-naive-layout} illustrates the sort order for a dataset with two attributes.

\TheSystem then sorts the data by this traversal. In other words, points in cell 0 (sorted by their sort dimension) come first, followed by cell 1, etc. Ties are broken arbitrarily.

\subsection{Basic Operation}
\label{sec:design:queryflow}

\TheSystem receives as input a filter predicate consisting of ranges over one or more attributes, joined by \text{{\small AND}}s. The intersection of these ranges defines a hyper-rectangle, and \TheSystem's goal is to find and process exactly the points within this hyper-rectangle (e.g., by aggregating them). At a high level, \TheSystem executes the following workflow (\Figure{query-flow}):
\begin{CompactEnumerate}
\item \textbf{Projection}: Identify the cells in the grid layout that intersect with the predicate's hyper-rectangle. For each such cell, identify the range of positions in storage, i.e. the \emph{physical index range}, that contains that cell's points (\Section{design:queryflow:vix-ranges}).
\item \textbf{Refinement: If applicable, take advantage of the ordering of points within each cell to shorten (or \emph{refine}) each physical index range that must be scanned (\Section{design:rectification}).}
\item \textbf{Scan}: For each refined physical index range, scan and process the records that match the filter.
\end{CompactEnumerate}

\subsubsection{Projection}
\label{sec:design:queryflow:vix-ranges}
In order to determine which points match a filter, \TheSystem first determines which cells contain the matching points.
Since the query defines a ``hyper-rectangle'' in the \mbox{$(d-1)$-dimensional} grid, computing intersections is straightforward. Suppose that each filter in the query is a range of the form $[q^s_i, q^e_i]$ for each indexed dimension $i$. If an indexed dimension is not present in the query, we simply take the start and end points of the range to be $-\infty$ and $+\infty$, respectively. Conversely, if the query includes a dimension not in the index, that filter is ignored at this stage of query processing.

\begin{figure}[t!]
\centering
\includegraphics[width=\columnwidth]{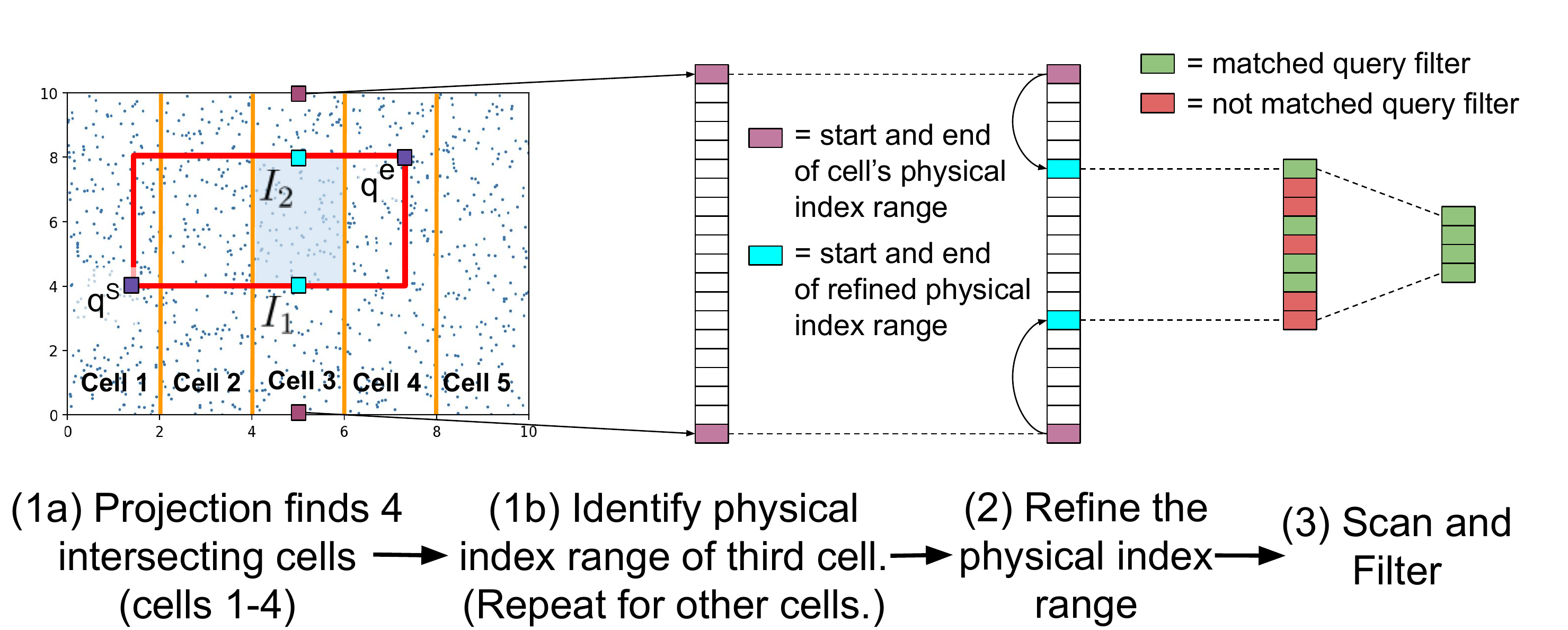}
\caption{Basic flow of \TheSystem's operation}
\label{fig:query-flow}
\end{figure}

The ``lower-left'' corner of the hyper-rectangle is $q^s = (q^s_0, \ldots, q^s_{d-1})$ and likewise for the ``upper-right'' corner $q^e$. Both are shown in \Figure{query-flow}. Then, we define the set of \emph{intersecting cells} as $
\{C_i \mid \mbox{cell}(q^s)_i \leq C_i \leq \mbox{cell}(q^e)_i \}$. \TheSystem keeps a \emph{cell table} which records the physical index of the first point in each cell.
Knowing the intersecting cells then easily translates to a set of physical index ranges to scan. 

\subsubsection{Refinement}
\label{sec:design:rectification}

When the query includes a filter over the sort dimension, \TheSystem uses the fact that points in each cell are ordered by the sort dimension to further refine the physical index ranges to scan. In particular, suppose the query includes a filter over the sort dimension $R.S$ of the form $a \leq R.S \leq b$. For each cell, \TheSystem finds the physical indices of both the first point $I_1$ having $R.S \geq a$ and the last point $I_2$ such that $R.S \leq b$. This narrows the physical index range for that cell down to $[I_1, I_2]$. The simplest way to find $[I_1, I_2]$ is by performing binary search within $C$ on the values in the sort dimension. This is possible only because the points in $C$ are stored contiguously in sorted order by the sort dimension. We discuss a faster way to refine, using models, in \Section{flattening:estimate}. If the query does not filter over the sort dimension, \TheSystem skips the refinement step.

\section{Optimizing the Grid}
\label{sec:opt}

\TheSystem's grid layout has several parameters that can be tuned, namely the number of columns allocated to each of the $d-1$ dimensions that form the grid, and which dimension to use as the sort dimension. Adjusting these parameters is the key way in which \TheSystem optimizes performance on a given query workload. We found that the ordering of the $d-1$ grid dimensions did not significantly impact performance.

\begin{figure}[!t]
\vspace{-1em}
    \centering
    \subfloat{
        \includegraphics[width=0.4\columnwidth,clip]{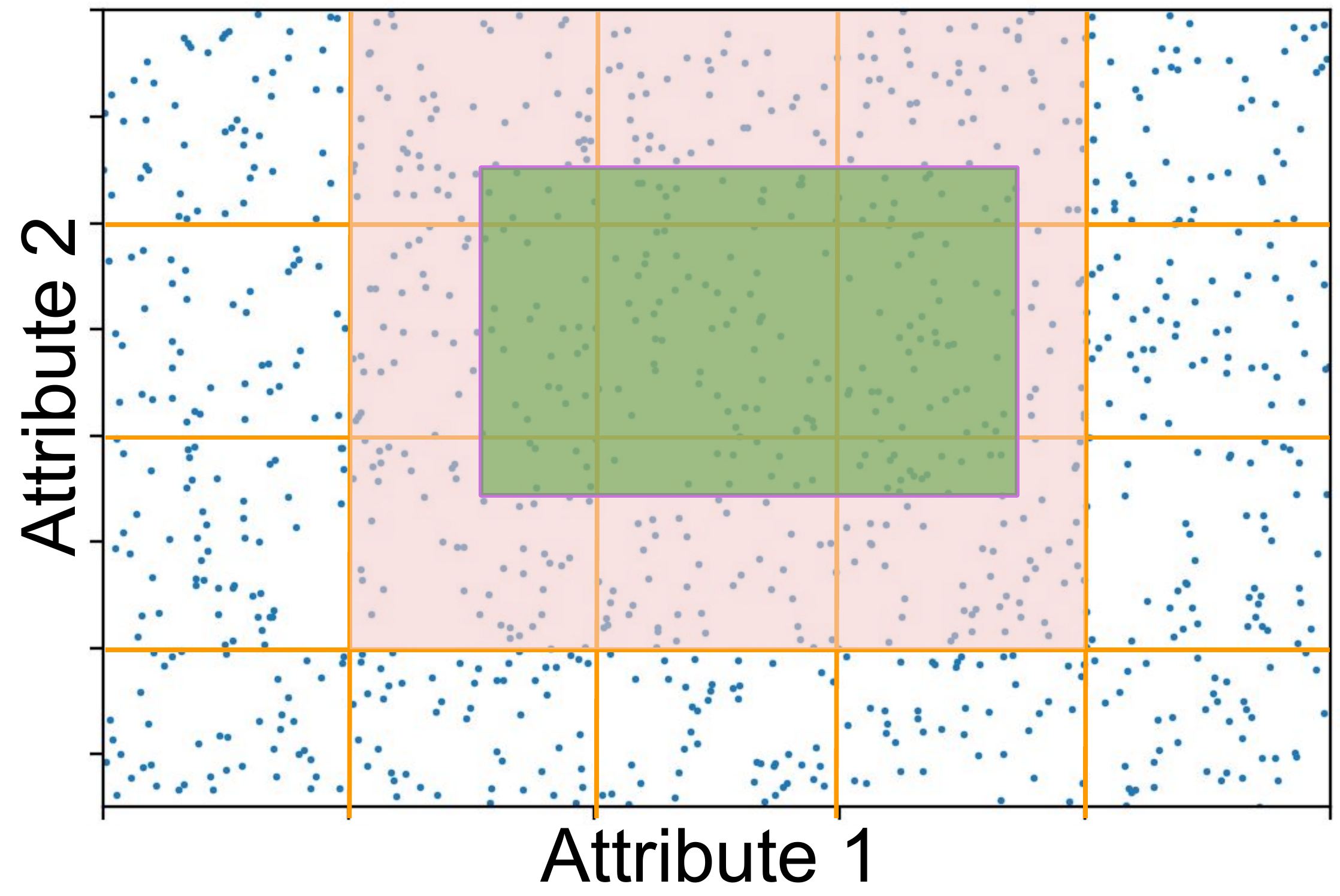}
        }
    \subfloat{
        \includegraphics[width=0.4\columnwidth,clip]{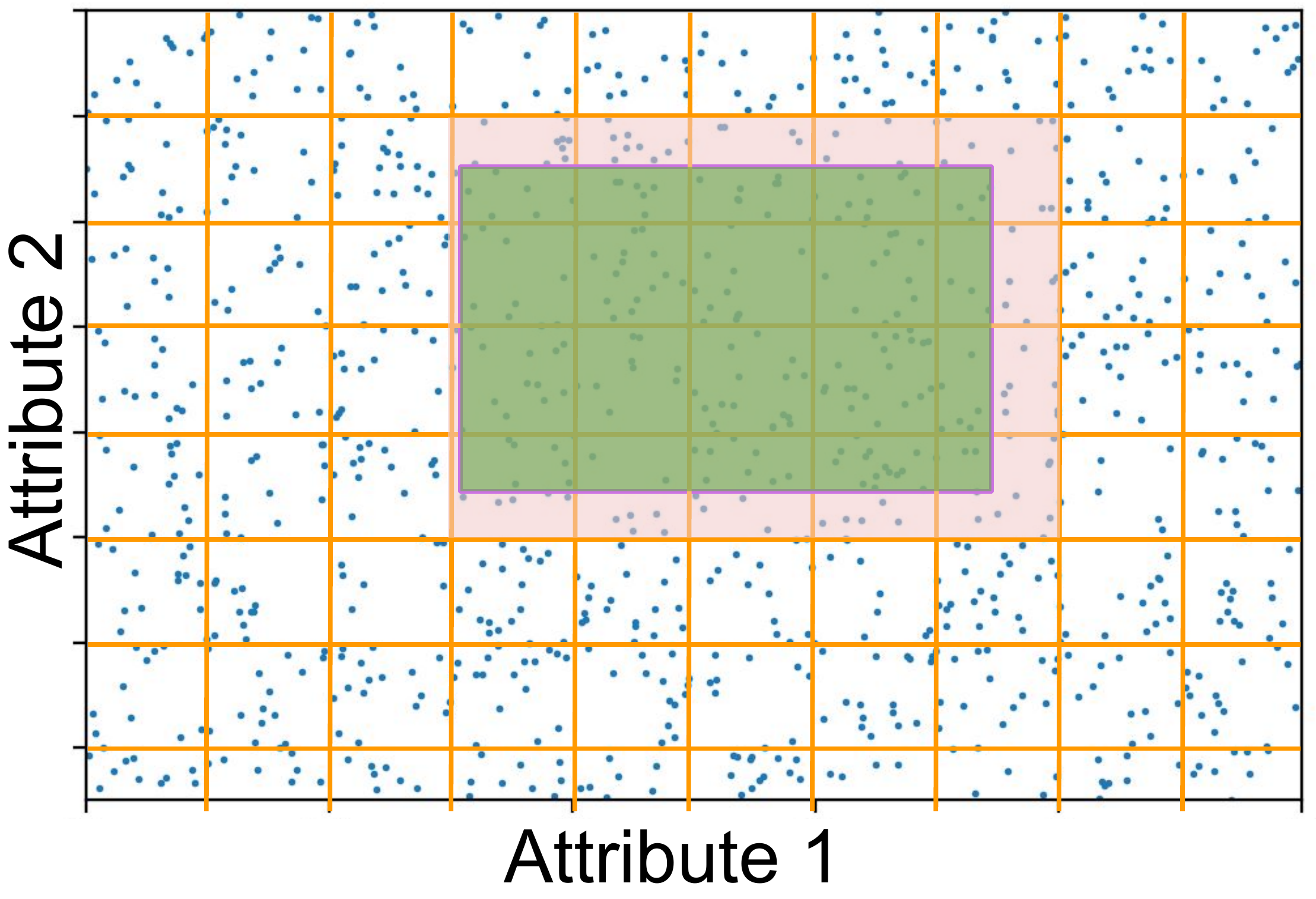}
        }
\caption{Doubling the number of columns  can increase the number of visited cells but decreases the number of scanned points that don't match the filter (light red).}
\vspace{-1em}
\label{fig:2d-naive-cells}
\end{figure}

Adding more columns in each dimension allows 
\TheSystem
to scan a rectangle that more tightly bounds the true query filter, which reduces the number of points that must be scanned
(\Figure{2d-naive-cells}). However, adding more columns also increases the number of sub-ranges, which incurs extra cost for projection and refinement.
Striking the right balance requires choosing a layout with an optimal number of columns in each dimension.

\TheSystem can also select the sort dimension. The sort dimension is special because it will incur no scan overhead; given a query, \TheSystem finds the precise sub-ranges to scan in the refinement step, so that the values in the sort dimension for scanned points are guaranteed to lie in the desired range. On the other hand, the grid dimensions do incur scan overhead because a certain column might only lie partially within the query rectangle. Therefore, the choice of sort dimension can have a significant impact on performance.

It is hard to select the optimal number of columns in each dimension because it depends on many interacting factors, including the frequency of queries filtering on that dimension, the average and variance of filter selectivities on that dimension, and correlations with other dimensions in both the data and query workload. The optimal sort dimension is also hard to select for similar reasons. Therefore, we optimize layout parameters using a cost model based approach. We first describe the cost model, then present the procedure that \TheSystem uses to optimize the layout.

\subsection{Cost Model}
Define a layout over $d$ dimensions as $L = (O, \{c_i\}_{0\le i < d-1})$, where $O$ is an ordering of the $d$ dimensions, in which the $d$th dimension is the sort dimension and $\{c_i\}_{0\le i < d-1}$ is the number of columns in the remaining $d-1$ grid dimensions.

Given a dataset $D$ and a layout $L$, we model the query time of any query $q$ as a sum of three parts, which correspond to the steps of the query flow from \Section{design:queryflow}. Each part consists of some measurable statistic $N$, which is multiplied by a variable weight $w$ which is a function of the dataset $D$, query $q$, and layout $L$, to produce an estimate of time taken on that step:
\begin{CompactEnumerate}
\item \textbf{Projection} contributes $w_pN_c$ to the query time, where $N_c$ is the number of cells that fall within the query rectangle, and $w_p$ is the average time to perform projection on a single cell. The weight $w_p$ is not constant across all datasets, queries, and layouts. For example, it is faster to identify a block of cells along a single grid dimension, which are adjacent on linear storage media, than a hypercube of cells along multiple grid dimensions which are non-adjacent.
\item \textbf{Refinement} contributes $w_rN_c$ to the query time, where $w_r$ is the average time to perform refinement on a cell. If the query $q$ does not filter on the sort dimension, refinement is skipped and $w_r$ is zero. Also, $w_r$ is lower if the cell is smaller, because the piecewise linear CDF for that cell (explained in \Section{flattening:estimate}) is likely less complex and makes predictions more quickly.
\item \textbf{Scan} contributes $w_sN_s$ to the query time, where $N_s$ is the number of scanned points, and $w_s$ is the average time to perform each scan. The weight $w_s$ depends on the number of dimensions filtered (fewer dimensions means fewer lookups for each scanned point), the run length of the scan (longer runs have better locality), and how many scans fall within exact sub-ranges (explained in \Section{eval:implementation}).
\end{CompactEnumerate} 

Putting everything together, our model for query time is:
\begin{equation}
Time(D, q, L) = w_pN_c + w_rN_c + w_sN_s
\label{eq:opt}
\end{equation}

Given a dataset $D$ and a workload of queries $\{q_i\}$, we find the layout $L$ that minimizes the average of \Equation{opt} for all $q \in \{q_i\}$.

\subsubsection{Calibrating the Cost Model Weights}
\label{sec:opt:calibration}
Since the four weight parameters $w = \{w_p, w_r, w_s\}$ vary based on the data, query and layout, \TheSystem uses models to predict $w$. The features of these weight models are statistics that can be measured when running the query on a dataset with a certain layout. These statistics include $N=\{N_c,N_s\}$, the total number of cells, the average, median, and tail quantiles of the sizes of the filterable cells, the number of dimensions filtered by the query, the average number of visited points in each cell, and the number of points visited in exact sub-ranges.

As we show in \Section{eval:index-creation}, the weight models are accurate across different datasets and query workloads. In particular, when new data arrives or the query distribution changes, \TheSystem needs only to evaluate the existing models, instead of training new ones. \TheSystem therefore trains the weight models once to calibrate to the underlying hardware. To produce training examples, \TheSystem uses an arbitrary dataset and query workload, which can be synthetic. \TheSystem generates random layouts by randomly selecting an ordering of the $d$ dimensions, then randomly selecting the number of columns in the grid dimensions to achieve a random target number of total cells. \TheSystem then runs the query workload on each layout, and measures the weights $w$ and aforementioned statistics for each query. Each query for each random layout will produce a single training example. In our evaluation, we found that 10 random layouts produces a sufficient number of training examples to create accurate models. \TheSystem then trains a random forest regression model to predict the weights based on the statistics.

One natural question to ask is whether a single random forest model can be trained to predict query time, instead of factoring the query time as weighted linear terms and training a model for each weight. However, a single model is inadequate because we want to accurately predict query times across a range of magnitudes; a model for query time would optimize for accuracy of slow queries at the detriment of fast queries. On the other hand, the weights span a relatively narrow range (e.g., the average time to scan a point will not vary across orders of magnitude), so are more amenable to our goal.

\subsubsection{Why Use Machine Learning?}
\label{sec:opt:why_ml}
\begin{figure}[]
    \centering
    \subfloat{
        \includegraphics[width=0.42\columnwidth,clip]{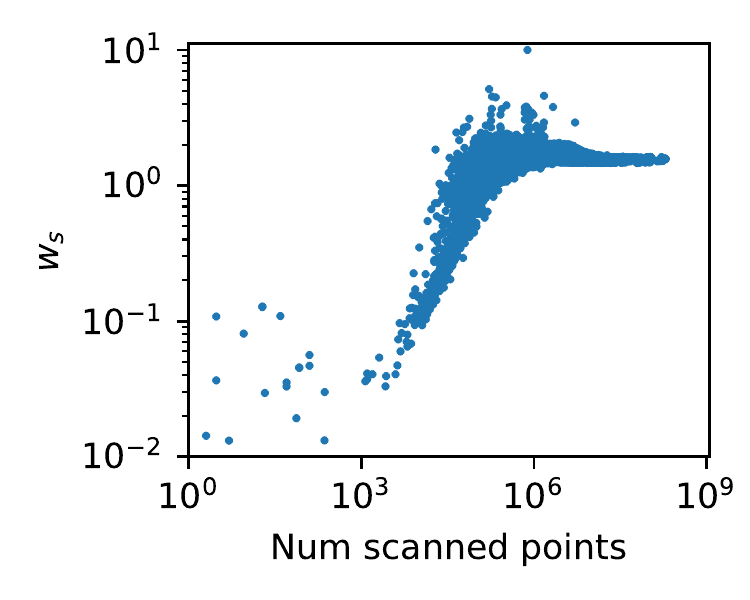}
        }
    \subfloat{
        \includegraphics[width=0.42\columnwidth,clip]{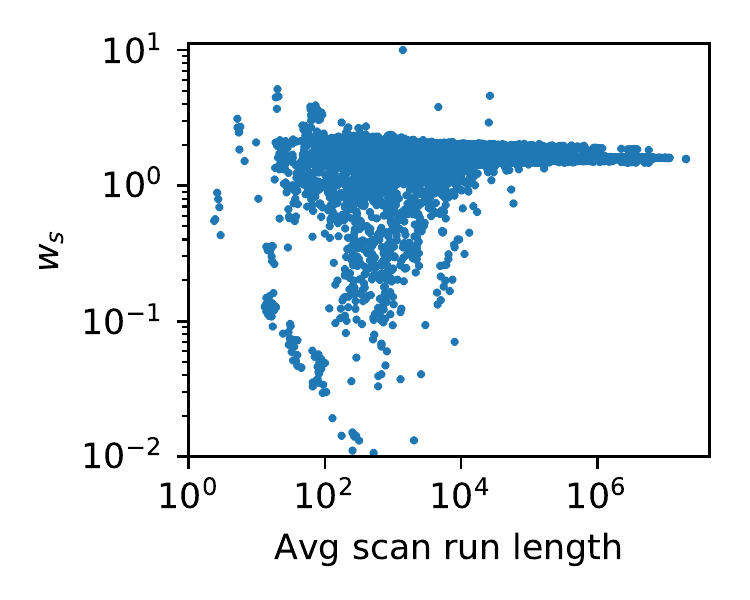}
        }
        \vspace{-0.2in}
\caption{$w_s$ (and by extension, scan time) is not constant and is difficult to model analytically because of its non-linear dependence on related features.}
\label{fig:scan_time}
\vspace{-0.2in}
\end{figure}

We model the cost using machine learning because query time is a function of many interdependent variables with potentially non-linear relationships that are difficult to model analytically. For instance, on 10k training examples, \Figure{scan_time} shows not only that the empirical average time to scan a point ($w_s$) is not constant, but also that its dependence on two related features (number of scanned points and average scan run length, which affects locality) is non-linear and does not follow an obvious pattern.

Indeed, we found that query time predicted using a simple analytical model that replaces the weight parameters of \Equation{opt} with fine-tuned constants has on average 9$\times$ larger difference from the true query time than our machine-learning based cost model. Furthermore, predicting the weight parameters using a linear regression model that uses the same input features as our random forest produces query time predictions with 4$\times$ larger difference from the true query time, which confirms that the features are interdependent and/or have non-linear relation with query time.

\subsection{Optimizing the Layout}
\label{sec:opt:training}
Given a calibrated cost model, \TheSystem optimizes its layout for a specific dataset and query workload as follows (pseudocode is provided in \Appendix{pseudocode}):
\begin{CompactEnumerate}
\item Sample the dataset and query workload, then flatten the data sample and workload sample using RMIs trained on each dimension.
\item Iteratively select each of the $d$ dimensions to be the sort dimension. Order the remaining $d-1$ dimensions that form the grid by the average selectivity on that dimension across all queries in the workload. This gives us $O$.
\item For each of these $d$ possible orderings, run a gradient descent search algorithm to find the optimal number of columns $\{c_i\}_{0\le i < d-1}$ for the $d-1$ grid dimensions. The objective function is \Equation{opt}. For each call to the cost model, \TheSystem computes the statistics $N=\{N_c,N_s\}$ and the input features of the weight models using the data sample instead of the full dataset $D$.
\item Select the layout with the lowest objective function cost amongst the $d$ layouts.
\end{CompactEnumerate}
Optimizing the layout is efficient (\Section{eval:index-creation}) because each iteration of gradient descent does not require building the layout, sorting the dataset, or running the query. Instead, statistics are either estimated using a sample of $D$ or computed exactly from the query rectangle and layout parameters.

\section{Learning from the Data}
\label{sec:flattening}
The simple index presented in \Section{design} does not consider or adapt to the underlying distribution of the data.
Here, we present two ways that \TheSystem learns its layout from the data. First, \emph{\TheSystem uses a model of each attribute to better determine column spacing.} Second, \emph{\TheSystem accelerates refinement within each cell using a model of the underlying data.}

\subsection{Flattening}
\label{sec:flattening:flattening}
The index in \Section{design} spaces columns equally, but this type of layout is inefficient when indexing highly skewed data: some grid cells will have a large number of points, causing \TheSystem to scan too many superfluous points.

If we were to have an accurate model of each attribute's distribution, i.e. its CDF, we could choose columns such that for each attribute, each column is responsible for approximately the same number of points. In practice, \TheSystem models each attribute using a Recursive Model Index (RMI), a hierarchy of models, e.g. linear models in our case, that is quick to evaluate~\cite{kraska}. The input to the model is the attribute value $v$; the output is the fraction of points with values $\leq v$. At query time, suppose that we would like to split the $k$th dimension into $n$ columns. A point with value $v$ in the $k$th dimension will be placed into column $\lfloor \text{CDF}(v) \cdot n \rfloor$. Since evaluating the RMI is efficient, we can efficiently determine the columns in each dimension that the query intersects.

\begin{figure}[t!]
\centering
\includegraphics[width=0.6\columnwidth]{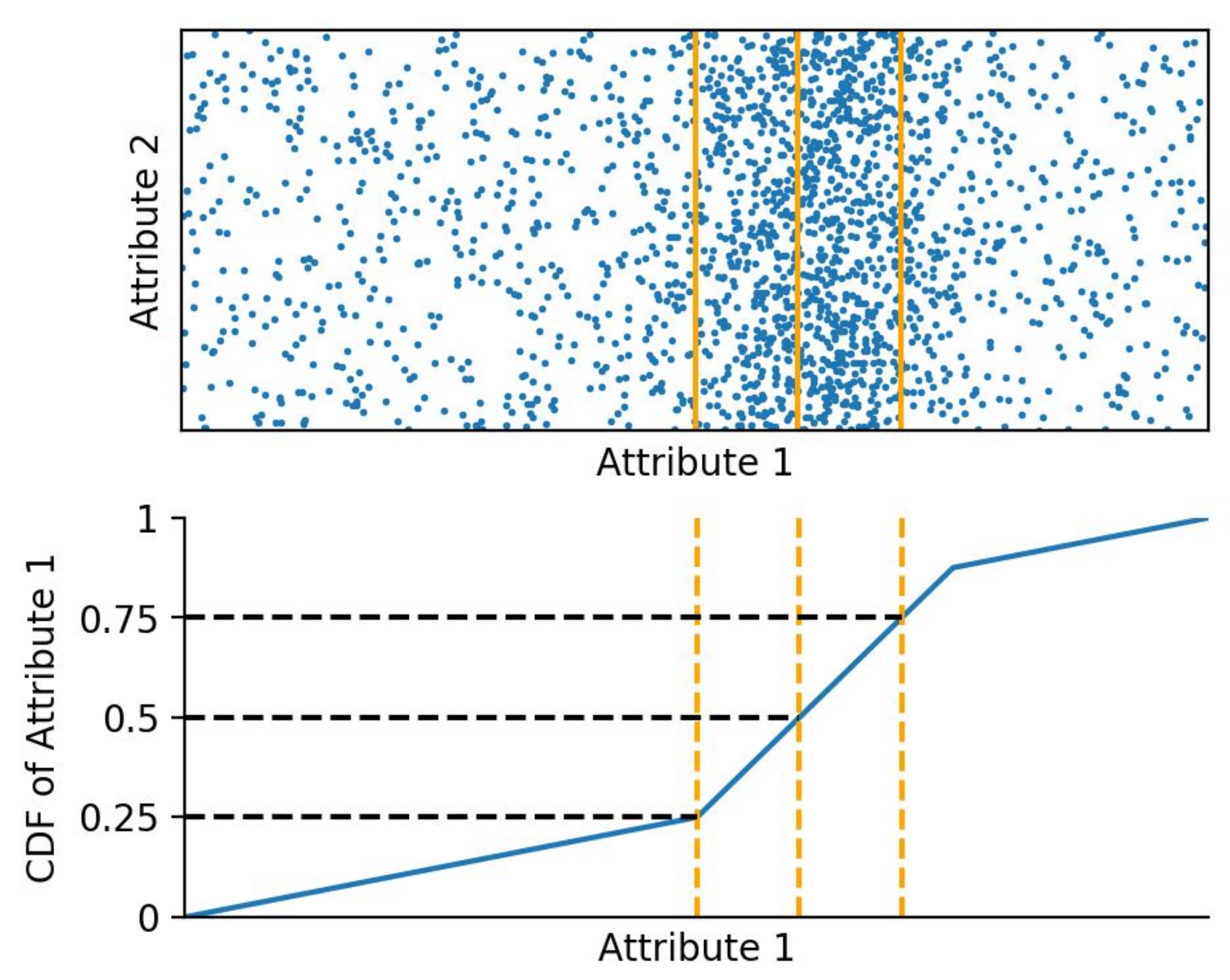}
\caption{By flattening, each of the four columns in a dimension will contain a fourth of the points.}
\label{fig:flattening}
\vspace{-0.2in}
\end{figure}

\Figure{flattening} shows example 2-D data and the result of applying this transformation to Attribute 1. The column boundaries are no longer equally spaced across the range of Attribute 1's values. Instead, they are equally spaced in terms of the CDF of Attribute 1. This means that each of the four columns has around \nicefrac{1}{4} of the points. Since the number of points in each column is evened out, we call this a \emph{flattened} layout. \TheSystem applies this flattening transformation for each grid dimension. Skewness is abundantly present in real-world data; on two of the datasets used in our evaluation (\Section{eval}), flattening provides a performance boost of $20 - 30\times$ over a non-flattened layout.

Note that while flattening may assign an equal number of points to each column of a single attribute, it does not guarantee that each cell in the final grid has a similar number of points. In particular, if two attributes are correlated, flattening each attribute independently will not yield uniformly sized cells. This may lead to some cells incurring a high scan overhead.
In practice, we found that modeling single attributes, i.e. assuming each dimension is independent, was sufficient. If necessary, adding more columns per dimension can further reduce the per-cell scan overhead. \TheSystem's layout training procedure (\Section{opt}) will choose the number of columns per dimension to trade off scan overhead with projection and refinement cost, mitigating the effect of non-uniform cell sizes. Additionally, if the correlation results in some cells being empty, those cells can be easily pruned using the cell table, incurring very little overhead.
However,  recent work \cite{correlationindex,correlationmaps} suggests that it might be possible to further reduce the index size by taking advantage of the correlation. Exploring such techniques for multi-dimensional clustered indexes remains future work. 

\subsection{Faster Refinement}
\label{sec:flattening:estimate}
The simple index from \Section{design:rectification} uses binary search over the sort dimension to refine the physical index range of each cell. In practice, since we may have to refine in every cell, binary search is too slow. Instead, \TheSystem builds a CDF model over the sort dimension values for each cell. \TheSystem uses a cell's model to estimate the endpoints of the refined physical index range, and then corrects any misprediction through a local search.

We want a model that can achieve a low average absolute error, while keeping the maximum error bounded to a reasonable value, in order for local search to be fast. Unfortunately, it is difficult to build an RMI with a target error bound. Instead, the model \TheSystem uses is a \emph{piecewise linear model} (PLM).

A PLM models a CDF by partitioning a sorted list of values $V$ into slices, each of which is modeled by a linear segment. Let $P(v)$ be the predicted index of value $v \in V$, determined by the segment responsible for the slice containing $v$, and let $D(v)$ be the index of the first occurrence of $v$. 
We require that the linear segments serve as a \emph{lower bound}
 on the true CDF values, i.e. $P(v) \leq D(v)$, with the property that for every segment, the average absolute error is less than a given threshold $\delta$:
\[ \frac{1}{|V|}\sum_{v \in V} |D(v) - P(v)| \leq \delta \]
The lower bound property allows us to turn this condition into: $\frac{1}{|V|}\sum_{v \in V} D(v) - P(v) \leq \delta$, which is much easier to achieve.

\TheSystem uses a greedy algorithm to partition $V$ into slices: for each $v \in V$ in increasing order, it adds $(v, D(v))$ to the segment for the current slice. If the segment's average error over the values in current slice exceeds $\delta$, it begins a new slice. The model records the smallest $v$ in each slice and forms a cache-optimized B-Tree over those values. At inference time, \TheSystem uses the B-Tree to find the appropriate segment and thus $P(v)$. The parameter $\delta$ encodes a tradeoff between size and speed (lower $\delta$ is faster): see \Section{eval:per-cell} for experiments tuning $\delta$ and a comparison of the PLM to other methods.
\section{Discussion}
\label{sec:discussion}
\NewPara{Tuning of traditional indexes.} Existing multi-dimensional indexes strive for fewer hyperparameters, typically only a page size, to lower the overhead of tuning.
However, fewer parameters also restricts the search space over which \TheSystem can optimize its layout, limiting the speedups it can achieve over existing indexes. Indeed, \Figure{pareto} demonstrates that simply tuning page size does not offer substantial performance improvement. In contrast, \TheSystem's grid layout intentionally offers a larger number of parameters over which to optimize, all of which can be automatically tuned by \TheSystem's learning procedure. 
As a result, \TheSystem can customize the layout for a particular query workload better than existing indexes.

\NewPara{Alternatives to grids.} \TheSystem is a learning-enhanced version of a basic grid index, but many alternative techniques to divide a multi-dimensional space exist (R-Tree, k-d tree, Z-order, etc.). 
We decided to use a grid structure for several reasons. First, it has a small space overhead: other multi-dimensional indexes use between 10MB and 1GB, but \TheSystem's grid uses less than 1kB (\Section{eval}), leaving ample room to add per-cell models. Second, the grid has low lookup latency, since it avoids pointer chasing. On the TPC-H dataset in \Section{eval}, \TheSystem with flattening takes 0.46ms to identify relevant grid cells (excluding refinement), while the k-d tree and hyperoctree take 8.9ms ($20\times$) and 1.8ms ($4\times$) to identify matching pages, respectively. This trend is consistent across the datasets we evaluate on. Z-order based indexes have low lookup times but expose no obvious parameters that can be tuned for the query workload.
Note that our flattening approach is necessary to keep scan times low by making sure the grid is not highly imbalanced.

\NewPara{Nearest Neighbor Queries.} Tree-based indexes that are used for geospatial data, such as k-d trees and R-trees, support $k-$nearest neighbor (kNN) queries. For example, a k-d tree locates the page with the query point and checks adjacent pages until all $k$ neighbors are found. \TheSystem can easily locate adjacent cells in its grid layout, allowing a similar kNN algorithm. However, since this paper does not focus on geospatial analytics, we exclude kNN queries from our evaluation.

\NewPara{Multi-dimensional CDFs.} In \Section{flattening:flattening}, we mentioned that correlated dimensions can yield non-uniform data after flattening. To address this issue, for each pair of correlated dimensions, one could train a 2-dimensional joint CDF, or train a conditional CDF that creates a 1-D model for attribute A within each column of attribute B. However, it is difficult to ensure that a multi-dimensional RMI model gives monotonic predictions along each dimension, which is a necessary property for partitioning points into columns; and conditional CDFs did not significantly improve performance in our benchmarks, but did significantly increase index size. Therefore, \TheSystem does not use multi-dimensional CDFs.
Efficiently modeling correlations between more dimensions is an active area of research~\cite{dim-sel, quick-sel}.
\section{Evaluation}
\label{sec:eval}
We first describe the experimental setup and then present the results of an
in-depth experimental study that compares \TheSystem with several other indexing methods on a variety of datasets and workloads. Overall, this evaluation shows that:
\begin{CompactEnumerate}
	\item \TheSystem achieves optimality \emph{across the board}: it is faster than, or on par with, every other index on the tested workloads. 
	However, the next best index changes depending on the dataset.
	On our datasets, \TheSystem is up to 187$\times$ faster than a single-dimensional clustered column index, up to 62$\times$ faster than a Grid File, up to 72$\times$ faster than a Z-order index, up to 250$\times$ faster than an UB-tree, up to 43$\times$ faster than a hyperoctree, and up to 48$\times$ faster than a k-d tree or R-tree.
    \item \TheSystem's index can take up to $50\times$ less space than the next fastest index.
	\item Even though we did not optimize \TheSystem for dynamic workloads, \TheSystem can train its layout and reorganize the records quickly for a new query distribution, typically in under a minute for a 300 million record dataset.
	\item \TheSystem's performance over baseline indexes improves with larger datasets and higher selectivity queries.
\end{CompactEnumerate}

\subsection{Implementation}
\label{sec:eval:implementation}
We implement \TheSystem in C++ on a custom column store that uses \emph{block-delta} compression: in each column, the data is divided into consecutive blocks of 128 values, and each value is encoded as the delta to the minimum value in its block. Our encoding scheme allows constant-time element access and is able to compress the datasets used in our evaluation by 77\%.

Our implementation uses 64-bit integer-valued attributes. Any string values are dictionary encoded prior to evaluation. Floating point values are typically limited to a fixed number of decimal points (e.g., 2 for price values). We scale all values by the smallest power of 10 that converts them to integers.

Our column store implementation has two other optimizations to improve scanning times:
\begin{CompactEnumerate}
\item If the range of data being scanned is \emph{exact}, i.e., we are guaranteed ahead of time that all elements within the range match the query filter, we skip checking each value against the query filter. For common aggregations, e.g. \texttt{\small COUNT}, this removes unnecessary accesses to the underlying data.
\item Similar to the idea of \cite{multirestrees}, our implementation allows indexes to speed up common aggregations like \texttt{\small SUM} by including a column in which the $i$th value is the cumulative aggregation of all elements up to index $i$. In the case of an exact range, the final aggregation result is simply the difference between the cumulative aggregations at the range endpoints.
Note that this is not a data cube as we can support arbitrary ranges instead of only pre-aggregated ranges.
\end{CompactEnumerate}
These additions are meant to demonstrate that \TheSystem can take advantage of features that existing indexes enjoy.
We show in \Section{eval:results:breakdown} that these optimizations are not required for \TheSystem: its performance benefits are due primarily to the optimality of the layout and not the details of the underlying implementation. Our random forest regression uses Python's Scipy library~\cite{scipy}.

To demonstrate that our column store implementation is comparable to the existing state of the art, we benchmark our column store with MonetDB, an open-source column store~\cite{monetdb}, by executing a query workload with full scans. Both MonetDB and our implementation were run single-threaded, with identical bit widths for each attribute, and without compression. Note that MonetDB does not support compression on numerical columns. Averaged over 150 aggregation queries on the TPC-H dataset (\Section{eval:results:real}), our scan times are within 5\% of MonetDB, showing that our column store implementation is on par with existing systems.

\vspace{-1em}
\subsection{Baselines}
\label{sec:eval:baselines}

We compare \TheSystem to several other approaches, each implemented on the same column store and using the same optimizations where applicable:
\begin{CompactEnumerate}
    \item \emph{Full Scan}: Every point is visited, but only the columns present in the query filter are accessed.
    \item \emph{Clustered Single-Dimensional Index}: Points are sorted by the most selective dimension in the query workload, and we learn a B-Tree over this sorted column using an RMI~\cite{kraska}. If a query filter contains this dimension, we locate the endpoints using the RMI. Otherwise, we perform a full scan. Since a clustered index spends a vast majority of its time scanning instead of indexing, an RMI-based approach performs comparably to a standard B-tree: their total query times are within 1\% of each other. Therefore, we show results only for the RMI-based index.
    \item \emph{Grid Files}~\cite{gridfile} index points by assigning them to cells in a grid, similar to \TheSystem. However, unlike \TheSystem, Grid File columns are determined incrementally and do not optimize for a query workload. Additionally, points in multiple adjacent cells may be stored together in the same bucket and are not sorted.
    We found that the Grid Files algorithm in \cite{gridfile} requires a long time to construct on heavily skewed data, so we omit results when it took over an hour.
    \item The \emph{Z-Order Index} is a multidimensional index that orders points by their \emph{Z-value}~\cite{zvalue}; contiguous chunks are grouped into pages.
    Given a query, the index finds the smallest and largest Z-value contained in the query rectangle and iterates through each page with Z-values in this range.
    \item The \emph{UB-tree}~\cite{ubtree} also indexes points using their Z-values. A query finds the range of points to scan in the same manner as the Z-Order Index. The UB-tree has the ability to skip forward to the next Z-value contained in the query rectangle, which avoids unnecessary scans.
    \item The \emph{Hyperoctree}~\cite{octree} recursively subdivides space equally into hyperoctants (the $d$-dimensional analog to 2-dimensional quadrants), until the number of points in each leaf is below a predefined but tunable page size.
    \item The \emph{k-d tree} recursively partitions space using the median value along each dimension, until the number of points in each leaf falls below the page size. The dimensions are selected in a round robin fashion, in order of selectivity.
    \item The \emph{R$^*$-Tree}\, is a read-optimized variant of the R-Tree that is bulk loaded to optimize for read query performance. We benchmark the R$^*$-Tree implementation from libspatialindex~\cite{libspatialindex}. On larger datasets, the R$^*$-Tree was prone to out-of-memory errors and was not included in benchmarks.
\end{CompactEnumerate}
While some techniques are also entirely read-optimized like \TheSystem (e.g., the Z-Order and $R^*$-Tree), others are inherently more write-friendly (e.g., UB-tree); we still include them for the sake of comparison while optimizing them for reads as much as possible (e.g., using dense cache-aligned pages).
Additional implementation details can be found in \Appendix{implementation}.

Our primary goal is to evaluate the performance of our multidimensional index as a fundamental building block for improving range request with predicates (e.g., filters) over one or more attributes. 
We therefore do not evaluate against other full-fledged database systems or queries with joins, group-bys, or other complex query operators.
While the impact of our multi-dimensional clustered index on a full query workload for an in-memory column-store database system would be interesting, it requires major changes to any available open-source column-store and is beyond the scope of this paper.
However, it should be noted that that even in its current form, \TheSystem could be directly used as a component to build useful services, such as a multi-dimensional key-value store. 

For a fair comparison, all benchmarks are implemented using a single thread without SIMD instructions.
We excluded multiple threads mainly because our baselines were not optimized for it. We discuss how \TheSystem could take advantage of parallelism and concurrency in \Section{future}.
All experiments are run on an Ubuntu Linux machine with an Intel Core i9 3.6GHz CPU and 64GB RAM.

\begin{table}[t!]
\small
\begin{tabular}{lllll}
\toprule
      & \textbf{sales} & \textbf{tpc-h}     & \textbf{osm}  & \textbf{perfmon}  \\
\midrule
\textbf{records}  & 30M & 300M & 105M & 230M \\
\textbf{queries}  & 1000 & 700 & 1000 & 800 \\
\textbf{dimensions}  & 6 & 7 & 6 & 6 \\
\textbf{size (GB)}  & 1.44 & 16.8 & 5.04 & 11   \\
\bottomrule
\end{tabular}
\centering
\caption{Dataset and query characteristics.}
\label{tab:data_params}
\vspace{-2em}
\end{table}

\begin{figure*}[t!]
    \subfloat{
        \includegraphics[width=0.27\textwidth,clip]{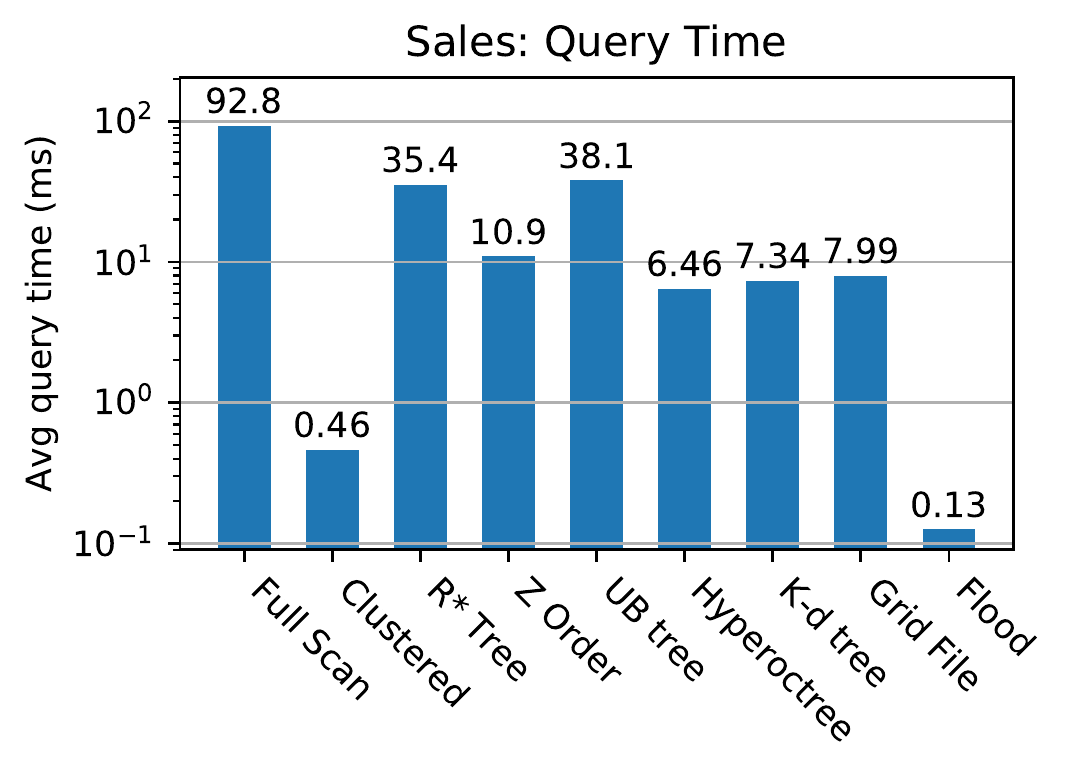}
        \label{fig:ds2_time}
    }
    ~
    \subfloat{
        \includegraphics[width=0.23\textwidth,clip]{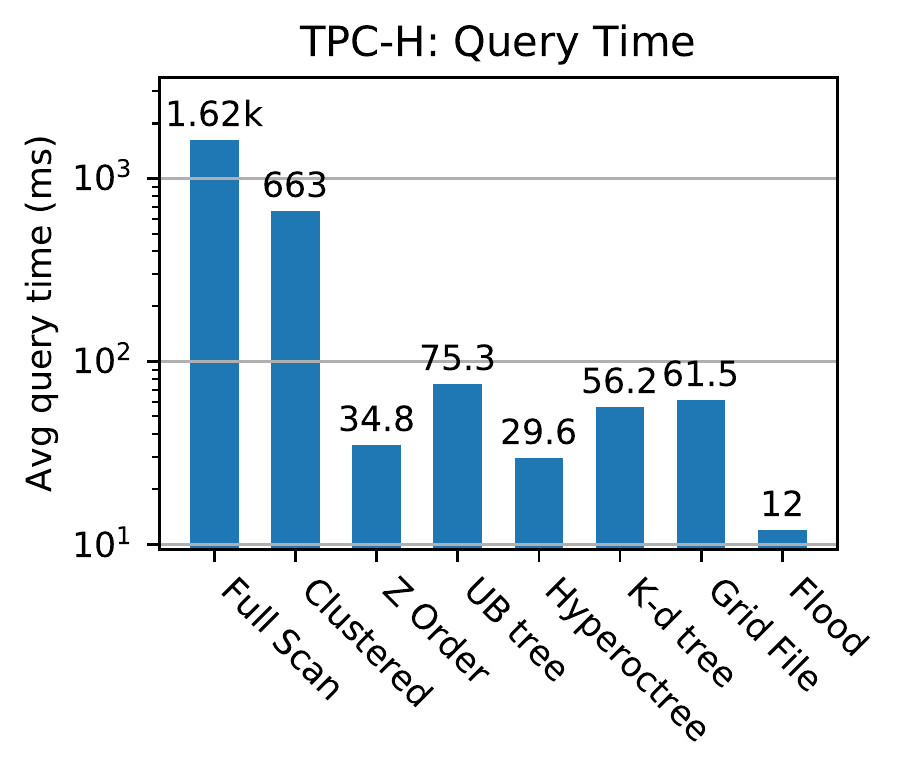}
        \label{fig:tpch_time}
        }
    ~
    \subfloat{
        \includegraphics[width=0.23\textwidth,clip]{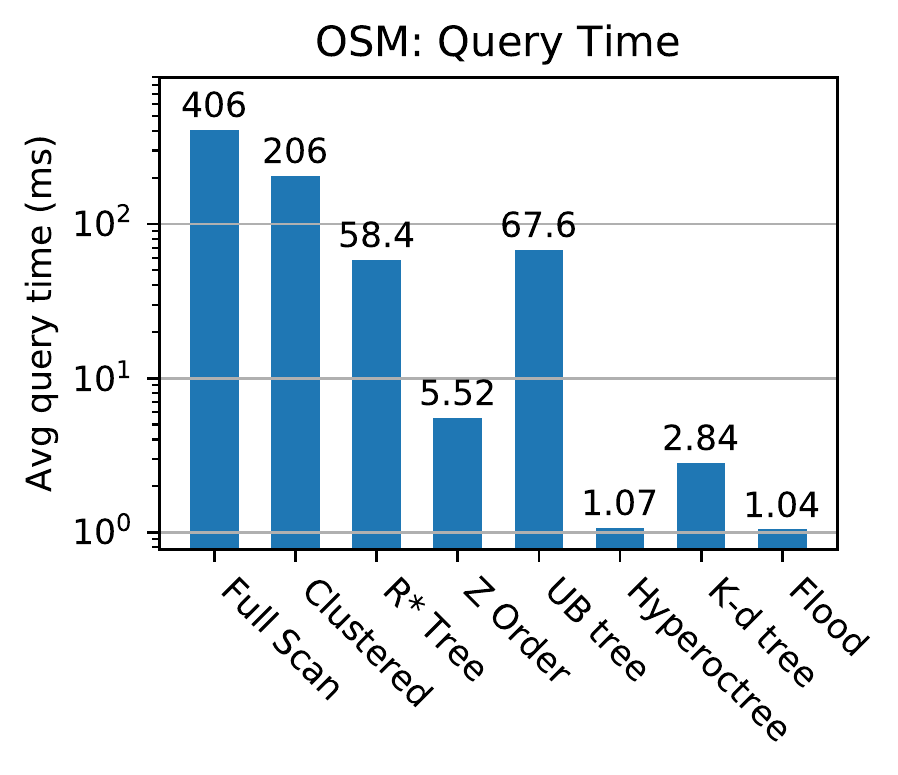}
        \label{fig:osm_time}
        }
    ~
    \subfloat{
        \includegraphics[width=0.23\textwidth,clip]{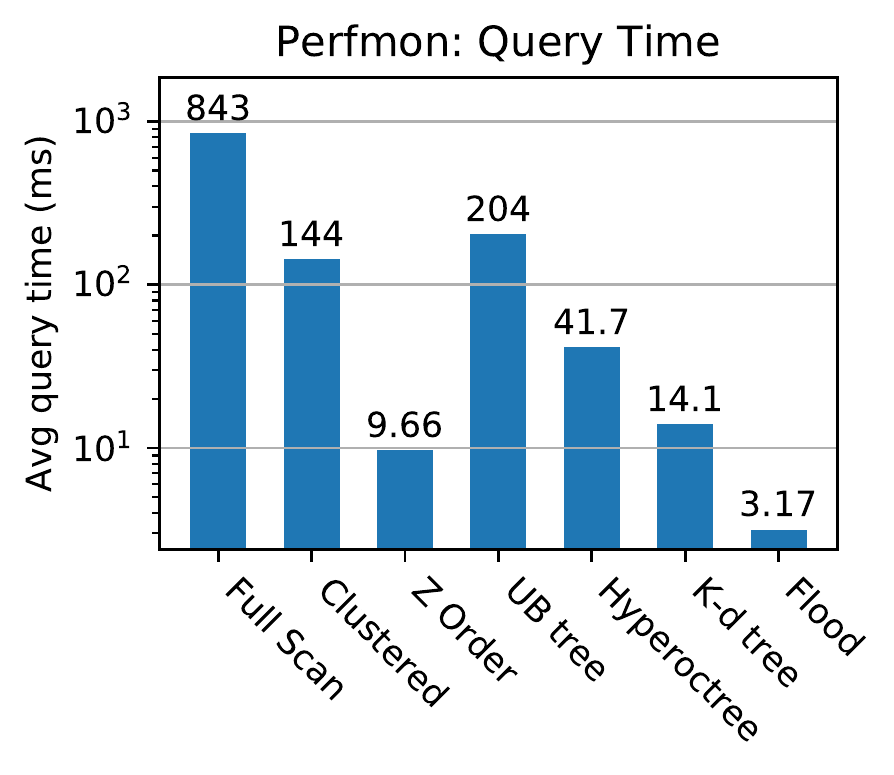}
        \label{fig:perfmon_time}
        }
    \vspace{-1em}
     \caption{
        Query speed of \TheSystem on all datasets. \TheSystem's index is trained automatically, while every other index is manually optimized for each workload to achieve the best performance. We excluded the R-tree for cases for which it ran out of memory. Note the log scale. 
        }
    \label{fig:punchline}
\end{figure*}

\begin{figure*}[t!]
    \centering
    \includegraphics[width=0.7\textwidth]
            {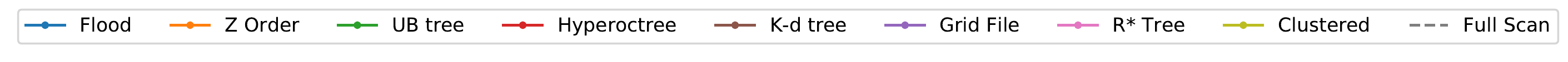}
    \\
    \vspace{-1em}
    \subfloat{
        \includegraphics[width=0.24\textwidth,clip]{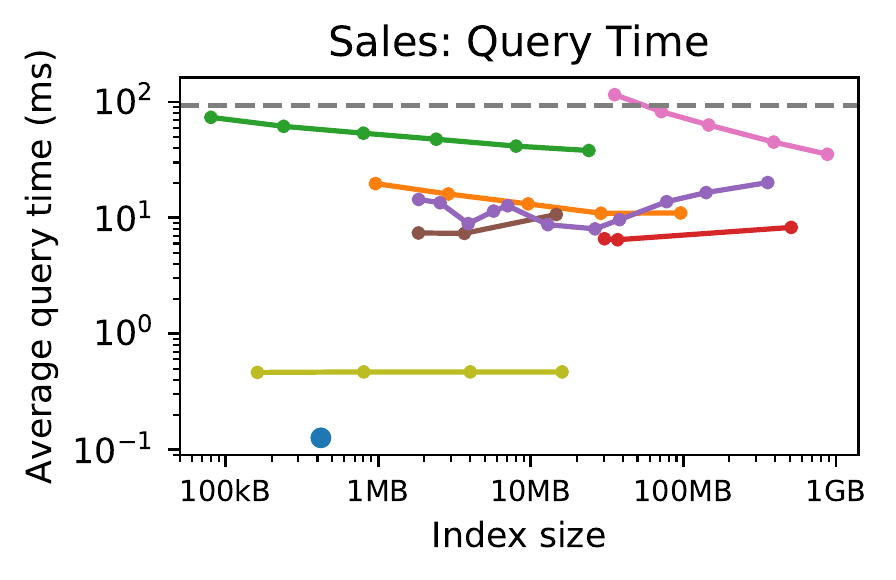}
        \label{fig:ds2_pareto}
        }
    ~
    \subfloat{
        \includegraphics[width=0.24\textwidth,clip]{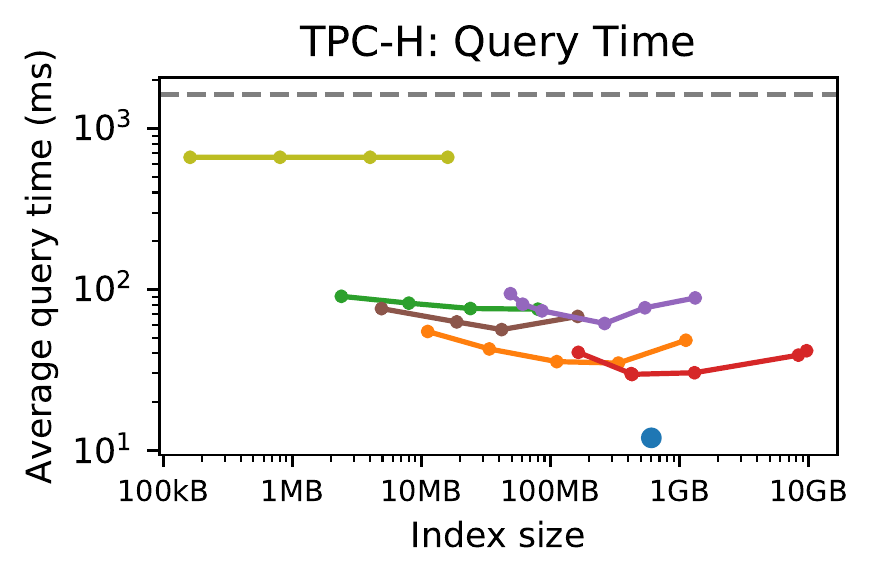}
        \label{fig:tpch_pareto}
        }
    ~
    \subfloat{
        \includegraphics[width=0.24\textwidth,clip]{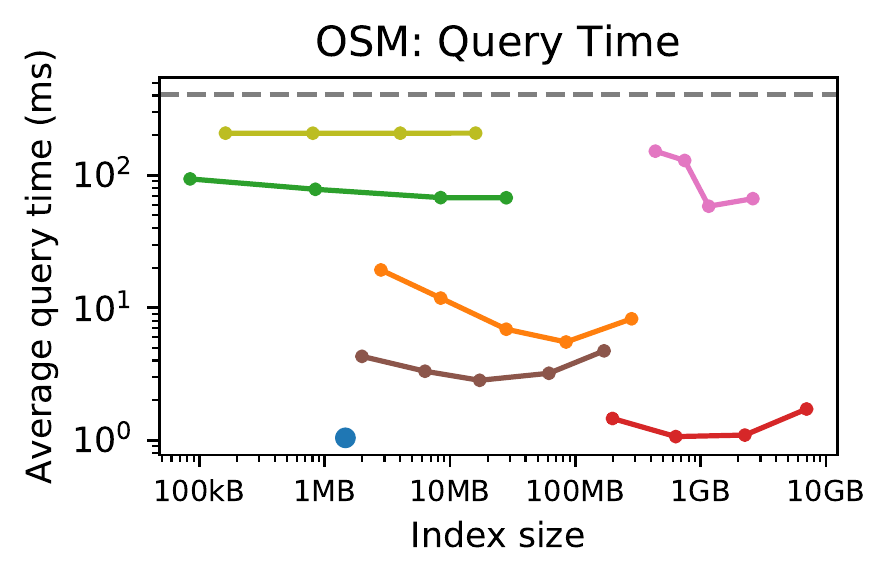}
        \label{fig:osm_pareto}
        }
    ~
    \subfloat{
        \includegraphics[width=0.24\textwidth,clip]{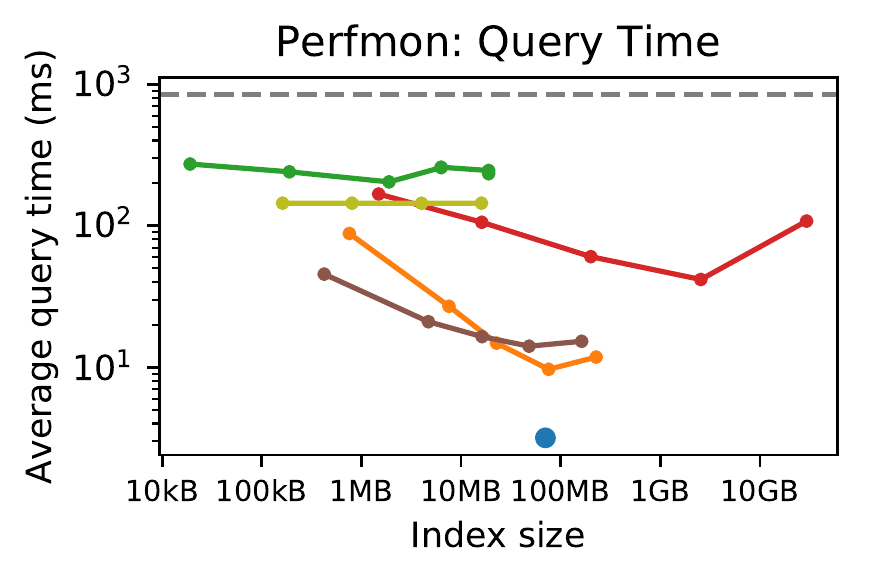}
        \label{fig:perfmon_pareto}
        }
    \vspace{-0.15in}
     \caption{
        \TheSystem (blue) sees faster performance with a smaller index, pushing the pareto frontier. Note the log scale.}
    \label{fig:pareto}
\end{figure*}

\subsection{Datasets}
\label{sec:eval:results:real}

We evaluate indexes on three real-world and one synthetic dataset, summarized in \Table{data_params}. Queries are either real workloads or synthesized for each dataset, and include a mix of range filters and equality filters.
The \textbf{Sales} dataset is a 6-attribute dataset and corresponding query workload drawn directly from a sales database at a commercial technology company. It was donated to us by a large corporation on the condition of anonymity. The dataset consists of 30 million records, with an anonymizing transformation applied to each dimension.
Each query in this workload was submitted by an analyst as part of report generation and analysis at the corporation.

Our second  real-world dataset, \textbf{OSM}, consists of all 105 million records catalogued by the OpenStreetMap~\cite{open-street-maps} project in the US Northeast. All elements contain 6 attributes, including an ID and timestamp, and 90\% of the records contain GPS coordinates.
Our queries answer relevant analytics questions, such as ``How many nodes were added to the database in a particular time interval?'' and ``How many buildings are in a given lat-lon rectangle?''
Queries use between 1 and 3 dimensions, with range filters on timestamp, latitude, and longtitude, and equality filters on type of record and landmark category. Each query is scaled so that the average selectivity is $0.1\% \pm 0.013\%$.

The performance monitoring dataset \textbf{Perfmon}  contains logs of all machines managed by a major US university over the course of a year. Our queries include filters over time, machine name, CPU usage, memory usage, swap usage, and load average.
The data in each dimension is non-uniform and often highly skewed. The original dataset has 23M records, but we use a scaled dataset with 230M records.

Our last dataset is \textbf{TPC-H}~\cite{tpch}.
For our evaluation, we use only the fact table, \texttt{lineitem}, with 300M records (scale factor 50) and create queries by using filters commonly found in the TPC-H query workload, with filter ranges scaled so that the average query selectivity is 0.1\%. Our queries include filters over ship date, receipt date, quantity, discount, order key, and supplier key, and either perform a \texttt{SUM} or \texttt{COUNT} aggregation. 

For each dataset, we generate a train and test query workload from the same distribution. \TheSystem's layout is optimized on the training set, and we only report results on the test set.

\subsection{Results}

\begin{figure*}[t!]
    \centering
    \includegraphics[width=0.52\textwidth]
            {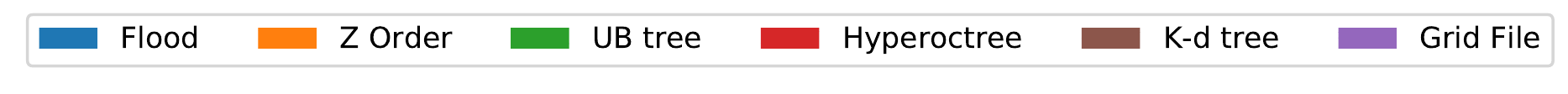} \\
    \vspace{-1.5em}
    \subfloat{
        \includegraphics[width=0.42\textwidth,clip]{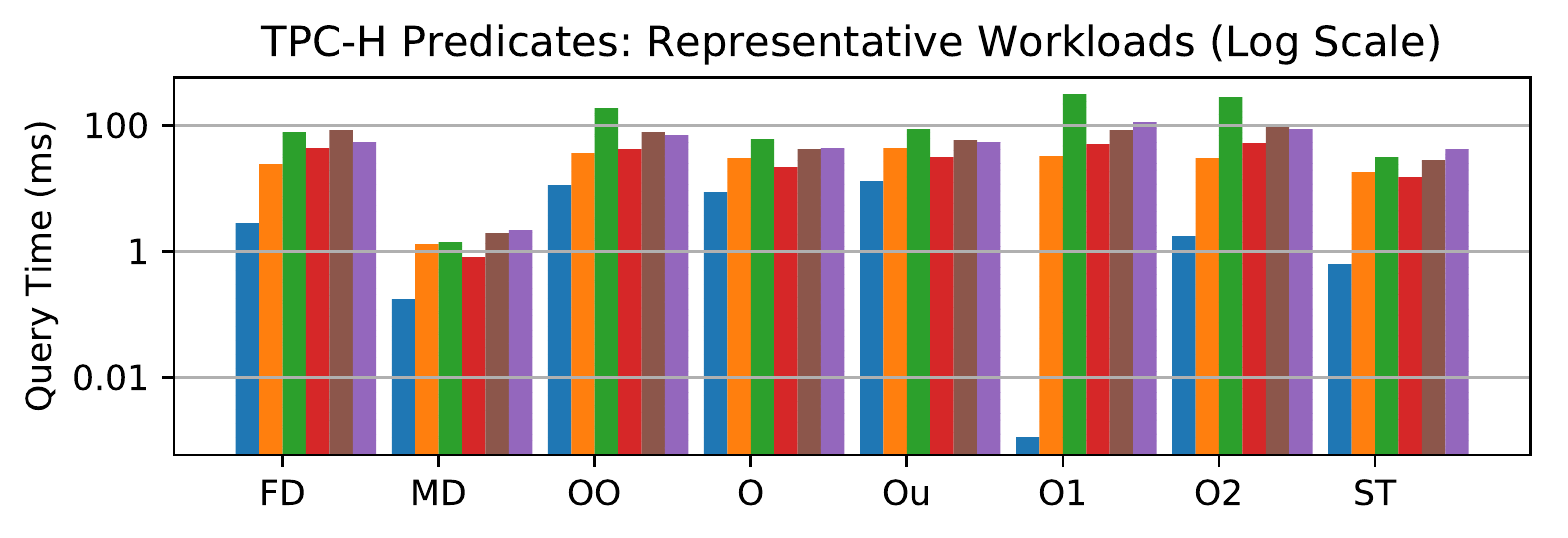}
        \label{fig:tpch_non_temporal}
        }
    ~
    \subfloat{
        \includegraphics[width=0.42\textwidth,clip]{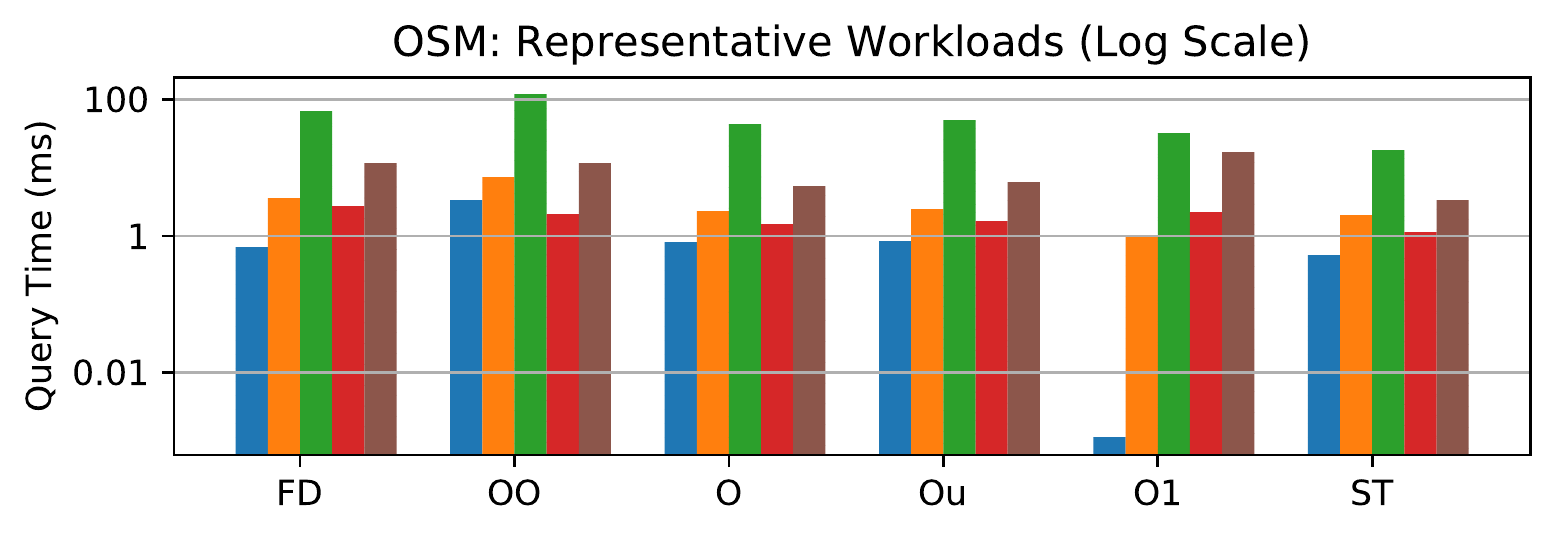}
        \label{fig:osm_non_temporal}
        }
    \vspace{-0.1in}
     \caption{\TheSystem and other indexes on workloads that have: fewer dimensions than the index (FD), as many dimensions as the index (MD), a skewed OLAP workload (O), a uniform OLAP workload (Ou), an OLTP workload over a single primary key (i.e., point lookups) (O1) and two keys (O2), a mixed OLTP $+$ OLAP workload (OO), and a single query type (ST). Note the log scale.
        }
    \label{fig:non-temporal}
    \vspace{-1em}
\end{figure*}

\begin{figure*}
    \centering
    \includegraphics[width=0.5\textwidth]
            {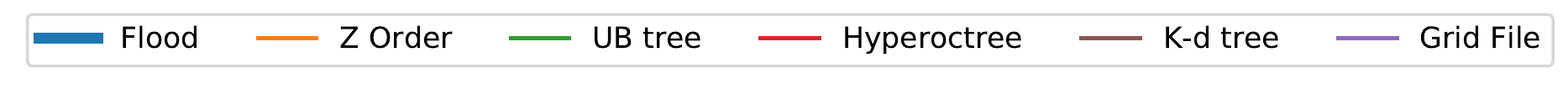}
            \vspace{-1em}
    \includegraphics[width=0.85\textwidth,clip]{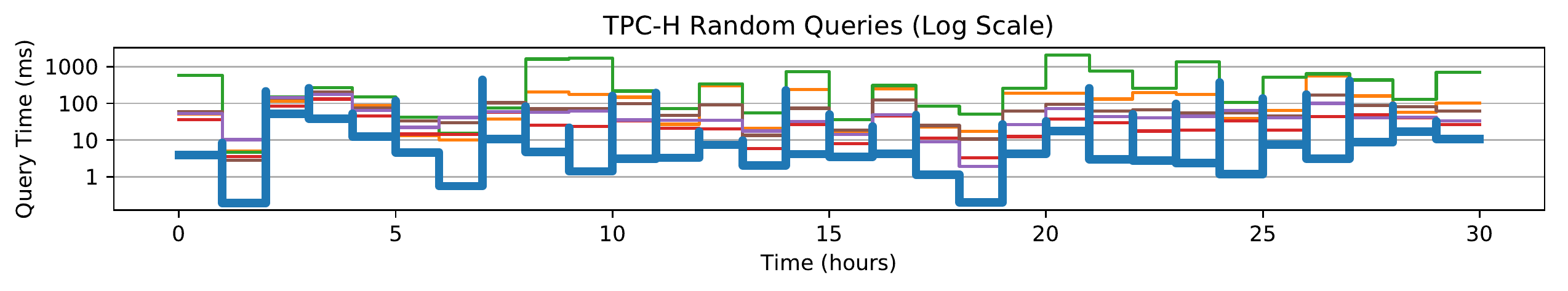}
        \vspace*{-5pt}
    \caption{\TheSystem vs. other indexes on 30 random query workloads, each for one hour. At the start of each hour, \TheSystem's performance degrades, since it is not trained for the new workload; however, it recovers in 5 minutes on average once the layout is re-learned, and beats the next best index by $5\times$ at the median. Note the log scale.}
    \vspace{-1em}
    \label{fig:temporal}
\end{figure*}

\NewPara{Overall Performance.} We first benchmark how well \TheSystem can optimize for a query workload compared to baseline indexes that are also optimized for the same query workload.
\Figure{punchline} shows the query time for each optimized index on each dataset. \TheSystem uses the layout learned using the algorithm in \Section{opt}, while we tuned the baseline approaches as much as possible per workload (e.g., ordered dimensions by selectivity and tuned the page sizes). 
This represents the best case scenario for the other indexes: that the database administrator had the time and ability to tune the index parameters.

On three of the datasets, \TheSystem achieves between $2.4\times$ and $3.3\times$ speedup on query time compared to the next closest index, and is always at least on-par, thus achieving the best performance {\em across-the-board}.
However, the next best system changes across datasets.
Thus, depending on the dataset and workload, \TheSystem can outperform each baseline by orders of magnitude.
For example, on the real-world sales dataset, \TheSystem is at least 43$\times$ faster than each multi-dimensional index, but only 3$\times$ faster than a clustered index. 
However, on the TPC-H dataset, \TheSystem is 187$\times$ faster than a clustered index.

On every dataset, \Figure{pareto} shows that \TheSystem beats the Pareto frontier set by the other multi-dimensional indexes. In particular, even though
\TheSystem's performance on OSM is on par with the hyperoctree, its index size is more than $20\times$ smaller. The hyperoctree thus has to spend much more memory for its performance than \TheSystem.
\TheSystem's space overhead comes partially from the grid layout metadata, but mostly (over 95\%) from the models of the sort attribute it maintains per cell.

\NewPara{Different Workload Characteristics.} 
In practical settings, it is unlikely that a database administrator will be able to manually tune the index for every workload change. The ability of \TheSystem to automatically configure its index for the current query workload is thus a significant advantage. 
We measure this advantage by tuning all indexes for the workloads in \Figure{punchline}, and then changing the query workload characteristics to:
\begin{CompactEnumerate}
    \item Single record filters, i.e. point lookups, using one or two ID attributes, as commonly found in OLTP systems.
    \item An OLAP workload, similar to the ones in \Figure{punchline}, that answer reasonable business questions about the underlying dataset. Some types of queries occur more often than others, skewing the workload.
    \item An OLAP workload where each query type is equally likely.
    \item An equal split of workloads (1) and (2), i.e., combined OLTP and OLAP queries.
    \item A workload with a single type of query, using the same dimensions with the same selectivities.
    \item A workload with fewer dimensions (a strict subset) than indexed by the baseline indexes.
\end{CompactEnumerate}

\Figure{non-temporal} shows the potential advantages \TheSystem can achieve over more static alternatives. 
\TheSystem consistently beats other indexes, though the magnitude of improvement depends on the dataset and query workload. For example, on TPC-H, \TheSystem achieves a speedup of more than $20\times$ on half the workloads, while on OSM, the median improvement is $2.2\times$.

\begin{table*}
\scriptsize
\begin{tabular}{lllll>{\columncolor[gray]{0.8}}l|llll>{\columncolor[gray]{0.8}}l|llll>{\columncolor[gray]{0.8}}l|llll>{\columncolor[gray]{0.8}}l}
\toprule
{} & \multicolumn{5}{l}{sales} & \multicolumn{5}{l}{tpc-h} & \multicolumn{5}{l}{osm} & \multicolumn{5}{l}{perfmon} \\
{} &      SO &    TPS &      ST &        IT &      TT &     SO &    TPS &       ST &        IT &      TT &       SO &    TPS &       ST &        IT &      TT &      SO &    TPS &      ST &       IT &      TT \\
\midrule
\textbf{Full Scan  } &  644 &   3.09 &  92.6 &  0 &  92.8 &  965 &   5.27 &  1580 &  0 &  1620 &  1090 &   3.83 &  403 &  0 &  406 &  990 &   3.52 &  833 &   0 &  843 \\
\textbf{Clustered  } &    3.18 &   3.09 &   0.462 &  6.76e-4 &   0.463 &  447 &   4.71 &   655 &  7.15e-4 &   662 &   478 &   4.50 &  207 &  8.92e-4 &  208 &  186 &   3.32 &  144 &   1.20e-3 &  144 \\
\textbf{Z Order    } &   57.9 &   4.00 &  10.9 &  0.0161 &  10.9 &   14.9 &   7.63 &    34.80 &  0.0267 &    34.8 &     6.85 &   8.37 &    5.5 &  0.0164 &    5.52 &    9.08 &   4.42 &    9.64 &   0.0146 &    9.66 \\
\textbf{UB tree    } &   55.7 &  14.5 &  38.0 &  0.0175 &  38.1 &   15.3 &  16.1 &    75.2 &  0.0284 &    75.3 &    22.5 &  31.3 &   67.5 &  0.0171 &   67.6 &   38.8 &  21.9 &  204 &   0.0120 &  204 \\
\textbf{Hyperoctree} &   38.8 &   3.34 &   6.11 &  0.353 &   6.46 &   20.8 &   4.38 &    27.8 &  1.77 &    29.6 &     2.36 &   3.59 &    0.812 &  0.253 &    1.07 &   33.8 &   3.47 &   28.2 &  13.4 &   41.7 \\
\textbf{K-d tree   } &   38.2 &   3.40 &   6.13 &  1.21 &   7.34 &   36.4 &   4.26 &    47.3 &  8.85 &    56.2 &     6.60 &   3.51 &    2.22 &  0.611 &    2.84 &   15.7 &   3.07 &   11.6 &   2.51 &   14.1 \\
\textbf{Grid File  } &   37.4 &   4.53 &   7.99 &  0.0594 &   7.99 &   36.9 &   5.28 &    59.5 &  1.88 &    61.5 &      N/A &    N/A &      N/A &       N/A &     N/A &     N/A &    N/A &     N/A &      N/A &     N/A \\
\textbf{Flood      } &    1.82 &   1.26 &   0.108 &  0.0182 &   0.128 &    5.90 &   5.53 &     9.96 &  2.02 &    12.0 &     3.13 &   2.39 &    0.717 &  0.328 &    1.05 &    4.26 &   2.77 &    2.84 &   0.327 &    3.17 \\
\bottomrule
\end{tabular}
\centering
\caption{Performance breakdown: scan overhead (SO), i.e. the ratio between points scanned and result size; average time spent scanning per scanned point, in nanoseconds (TPS); average time spent scanning, in milliseconds (ST); average time spent indexing (for \TheSystem this includes projection and refinement), in milliseconds (IT); total query time, in milliseconds (TT). SO$\times$TPS is proportional to ST, and ST+IT+$\epsilon$=TT (a small fraction of query time is spent neither scanning nor indexing). R$^*$-tree omitted because instrumentation for collecting statistics was inadequate in~\cite{libspatialindex}.}
\label{tab:punchline_analysis}
\vspace{-0.4in}
\end{table*}

\begin{figure*}[t!]
    \centering
    \subfloat{
        \includegraphics[width=0.23\textwidth,clip]{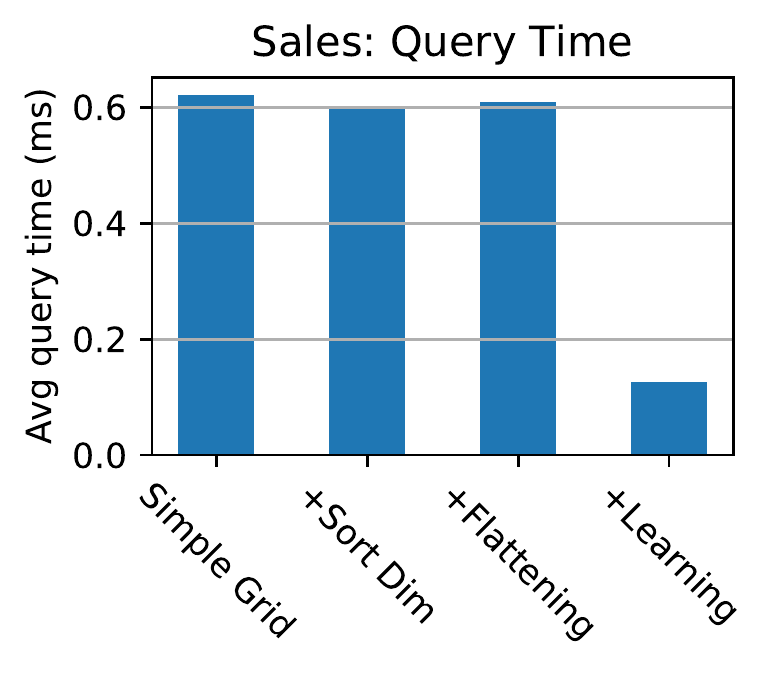}
        \label{fig:lesion_ds2}
        }
    ~
    \subfloat{
        \includegraphics[width=0.23\textwidth,clip]{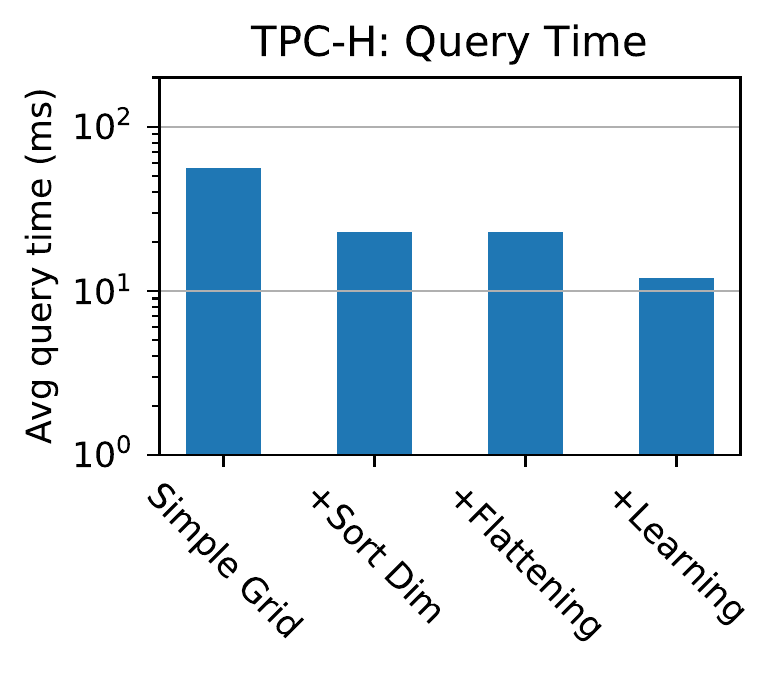}
        \label{fig:lesion_tpch}
        }
    ~
    \subfloat{
        \includegraphics[width=0.23\textwidth,clip]{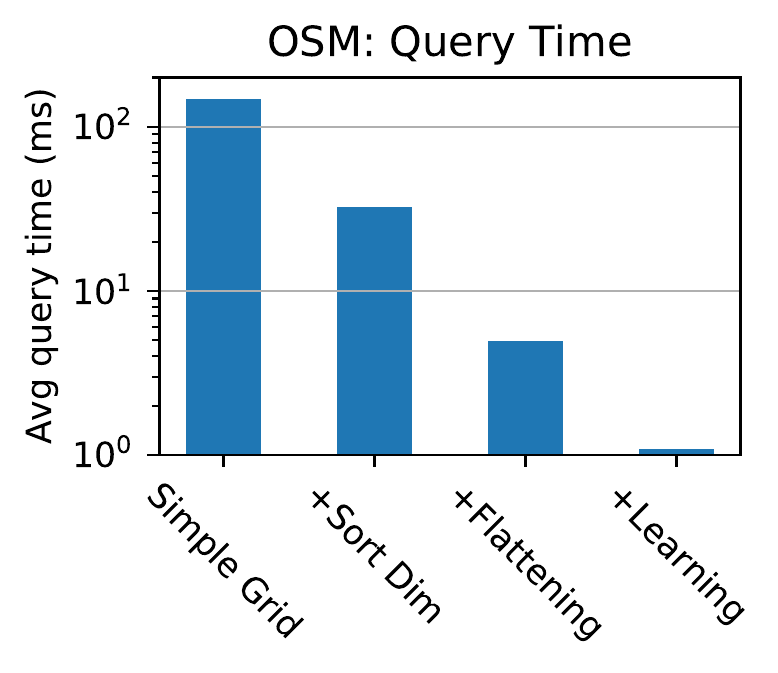}
        \label{fig:lesion_osm}
        }
    ~
    \subfloat{
        \includegraphics[width=0.23\textwidth,clip]{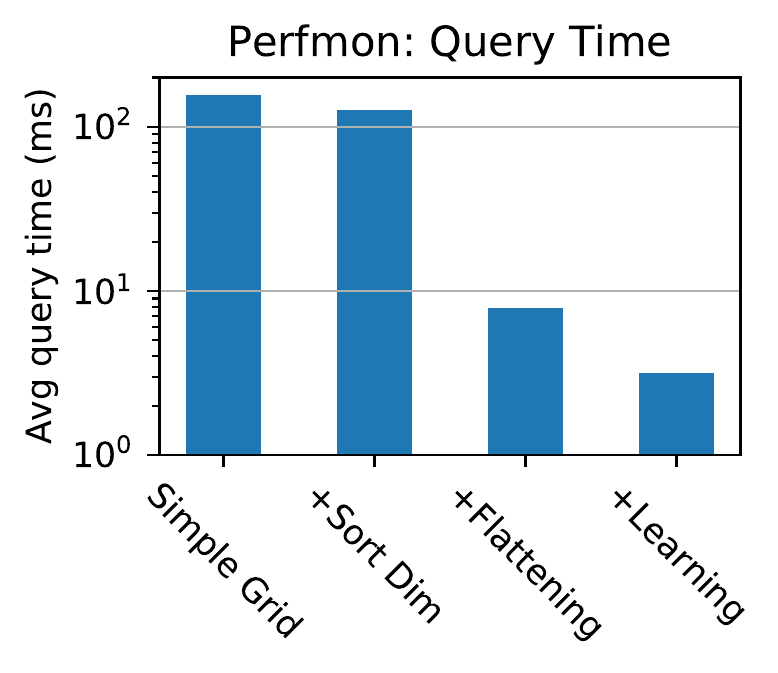}
        \label{fig:lesion_perfmon}
        }
    \vspace{-1em}
     \caption{
        Flattening and learning help \TheSystem achieve low query times but is workload dependent. 
        }
    \label{fig:lesion}
    \vspace{-0.1in}
\end{figure*}

\NewPara{Dynamic Query Workload Changes.}
Here, we demonstrate how the performance of \TheSystem varies over several random workloads, when the administrator does not tune the other indexes. 
We created 30 random workloads for the TPC-H dataset. 
Each workload runs for one hour and consists of at most 10 distinct query types, and each query type in turn consists of up to 6 dimensions, both chosen uniformly at random. The selectivities of each dimension are chosen randomly, with the constraint that all queries have an average total selectivity of around 0.1\% and are more selective on key attributes. 

\Figure{temporal} shows the results over time with \TheSystem being the only index that changes from one hour to the next (all others were kept fixed and tuned for the workload in \Figure{punchline}). At the start of each hour, a new query workload is introduced, and we trigger \TheSystem's retraining. During the retraining phase, which we assume happens on a separate instance, \TheSystem runs the new queries on its old layout, causing brief performance degradation and producing a spike at the start of each hour. It only switches to the new, more performant layout once retraining is finished. \TheSystem outperforms all other indexes, showing a median improvement of more than $5\times$ over the closest competitor, with 30\% of queries achieving more than a $10\times$ speedup. The results suggest that \TheSystem is able to generalize well by adapting to new and unforeseen workloads.

\Figure{temporal} also highlights the importance of learning from the query workload. When transitioning to the next query workload, \TheSystem's performance often worsens, since the current layout is usually not suitable for the new workload. Re-learning the layout based on the new workload lowers query time back lower than other indexes. Learning a layout is therefore (a) effective at adapting to new query workloads and (b) crucial to \TheSystem's performance improvement over other indexes. We leave the detection of workload changes to future work (\Section{future}).

While the results are encouraging, it is also important to consider the time it takes to adjust to a new query workload. \TheSystem takes at most around 1 minute to adapt to a new query workload, but it more than makes up for this adjustment period through improved performance on the subsequent workload. We evaluate index creation time in further detail in~\Section{eval:index-creation}.

\NewPara{Performance Breakdown.}
Where does \TheSystem's advantage over baseline indexes come from? We look at the \emph{scan overhead}, the ratio of total points scanned by the index to points matching the query. The scan overhead is implementation agnostic: it relies neither on the machine nor on the implementation of the underlying column store. A high scan overhead suggests that the index wastes time scanning unnecessary points. Since all indexes spend the vast majority of their time scanning, the scan overhead is a good proxy for overall query performance.

\Table{punchline_analysis} shows that \TheSystem achieves the lowest scan overhead (SO) on three out of four datasets
, which confirms that \TheSystem's optimized layout is able to better isolate records that match a query filter. Additionally, \TheSystem usually spends less time per scanned point (TPS) because \TheSystem avoids accessing the sort dimension. As a result, \TheSystem consistently achieves the lowest scan time (ST), which is proportional to the product of SO and TPS. This more than makes up for \TheSystem's higher index time (IT), which includes the time to project and refine.
Indexes based on Z-order incur the cost of computing Z-values and thus have a higher time per scanned point. Tree-based indexes have the highest index time due to the overhead of tree traversal.

Which of \TheSystem's components is responsible for its performance? \Figure{lesion} shows the incremental benefit of (1) sorting the last indexed dimension instead of creating a $d$-dimensional histogram, (2) flattening the data instead of using columns of fixed width, and (3) adapting to the query workload using the training procedure from \Section{opt}. The baseline system is a ``Simple Grid'' on all $d$ dimensions, with the number of columns in each dimension proportional to that dimension's selectivity.

Sorting by the last dimension offers marginal benefits: it allows more columns to be allocated to the first $d-1$ dimensions, increasing the resolution of the index along those dimensions without increasing the total number of cells. The biggest improvements come from flattening and learning the layout from the queries. However, the effect of each varies across datasets. Flattening benefits OSM and Perfmon since both datasets have heavily skewed attributes. Since Sales and TPC-H data are fairly uniform, using a non-flattened layout performs equally well. Finally, learning from queries provides major performance gains on all datasets, corroborating results from \Figure{temporal}.

\subsection{Scalability}
\label{sec:eval:results:breakdown}

\begin{figure}[]
    \centering
    \includegraphics[width=0.8\columnwidth]
            {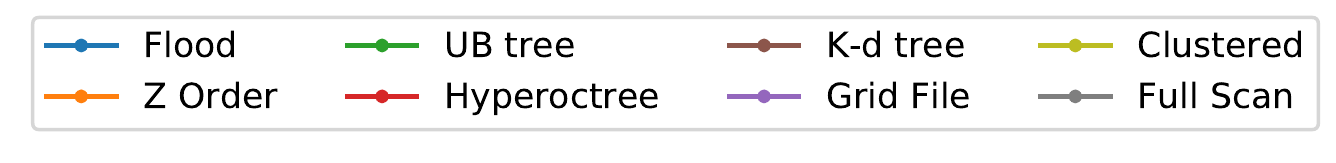}
    \\[-0.5ex] 
    \vspace{-0.1in}
    \subfloat{
        \includegraphics[width=0.48\columnwidth,clip]{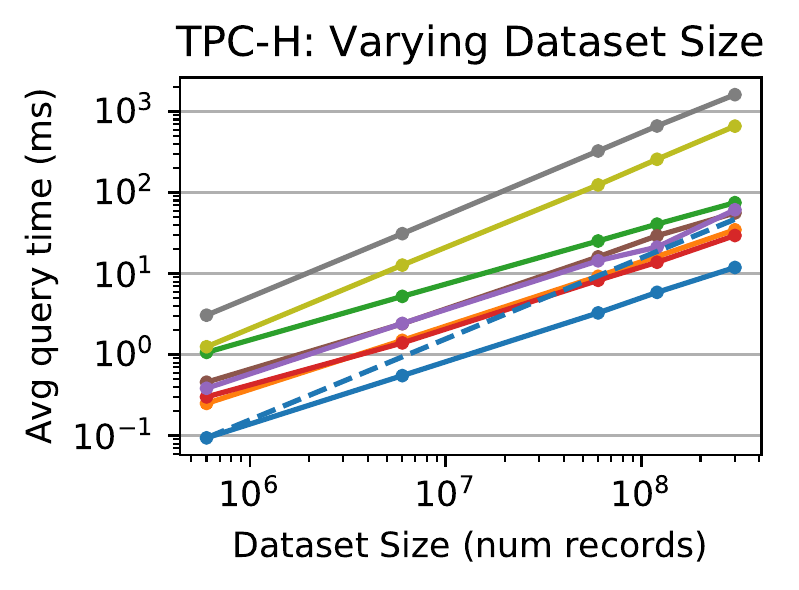}
        \label{fig:breakdown_dataset_size}
        }
    \subfloat{
        \includegraphics[width=0.48\columnwidth,clip]{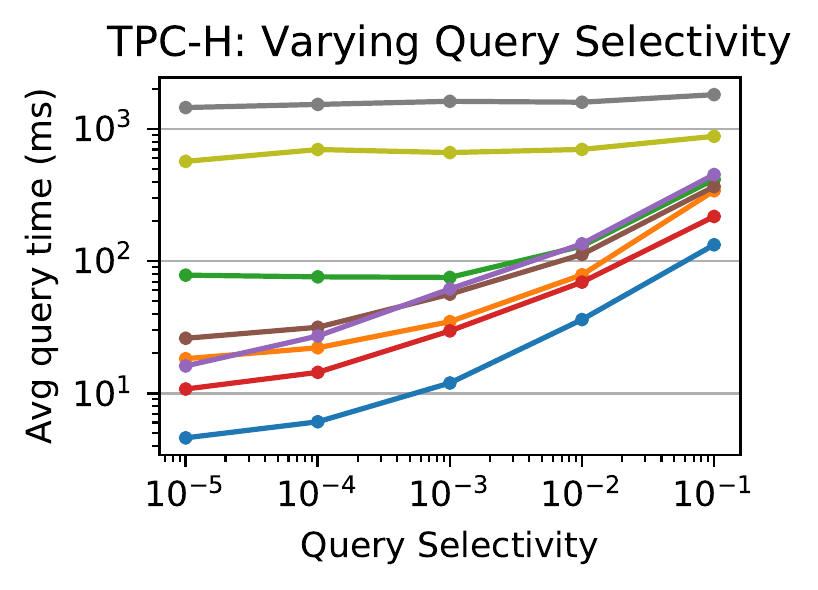}
        \label{fig:breakdown_selectivity}
        }
    \vspace{-0.1in}
     \caption{
        \TheSystem's performance scales both with dataset size and query selectivity. The dashed blue line depicts what linear scaling would like like.
        }
    \label{fig:breakdown}
    \vspace{-0.2in}
\end{figure}

\NewPara{Dataset Size.}
To show how \TheSystem scales with dataset size, we sample records from the TPCH dataset to create smaller datasets. We train and evaluate these smaller datasets with the same train and test workloads as the full dataset.
\Figure{breakdown_dataset_size} shows that the query time of \TheSystem grows sub-linearly. As the number of records grows, the layout learned by \TheSystem uses more columns in each dimension, which results in more cells. The extra overhead incurred by processing more cells is outweighed by the benefit of lowering scan overhead.

\NewPara{Query Selectivity.} To show how \TheSystem performs at different query selectivities, we scale the filter ranges of the queries in the original TPC-H workloads up and down equally in each dimension in order to achieve between 0.001\% and 10\% selectivity. \Figure{breakdown_selectivity} shows that \TheSystem performs well at all selectivities. The performance benefit of \TheSystem is less apparent at 10\% selectivity because all indexes are able to incur lower scan overhead when more points fall into the query rectangle.

\begin{figure}[]
    \centering
    \includegraphics[width=0.8\columnwidth]
            {figures/legend_scaling_dimensions.pdf}
    \\[-0.1in] 
    \vspace{-0.1in}
    \subfloat{
        \includegraphics[width=0.48\columnwidth,clip]{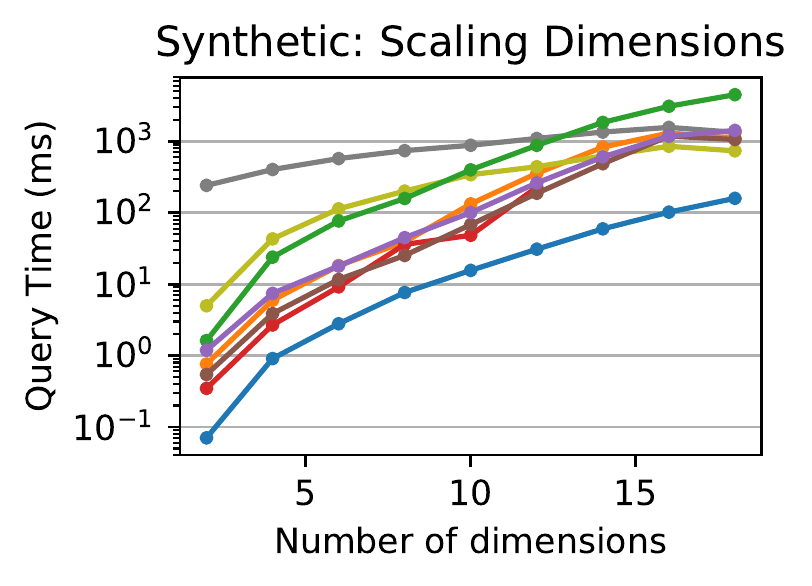}
        \label{fig:breakdown_dimensions}
        }
    \subfloat{
        \includegraphics[width=0.48\columnwidth,clip]{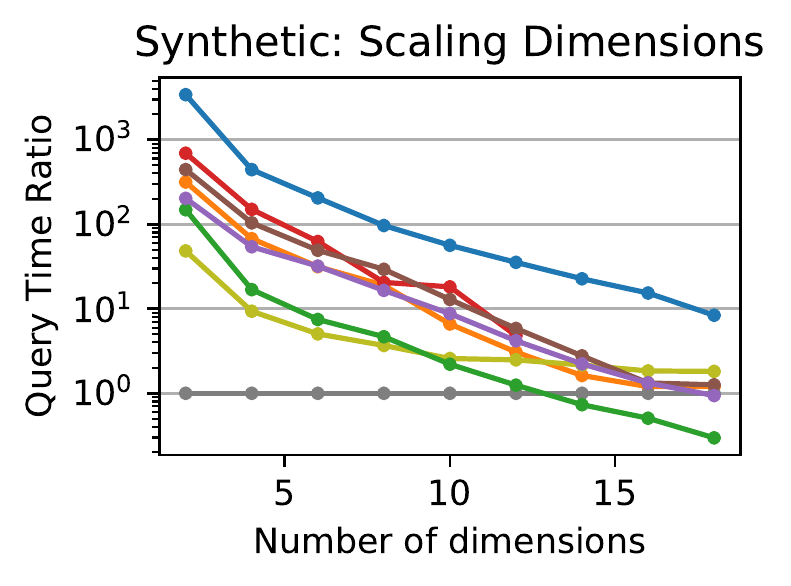}
        \label{fig:breakdown_dimensions_ratio}
        }
    \vspace{-0.1in}
     \caption{
        (a): Query time as number of dataset dimensions varies. (b): Ratio of the index's query time to the time for a full scan.
        }
    \label{fig:scaling_dimensions}
    \vspace{-0.2in}
\end{figure}

\NewPara{Number of Dimensions.}
To show how \TheSystem scales with dimensions, we create synthetic $d$-dimensional datasets $(d \leq 18)$ with 100 million records whose values in each dimension are distributed uniformly at random. For each dataset, we create a query workload of 1000 queries. The number of dimensions filtered in the queries varies uniformly from 1 to $d$.
If a query has a filter on $k$ dimensions, they are the first $k$ dimensions in the dataset.
For each query, the filter selectivity along each dimension is the same and is set so that the overall selectivity is 0.1\%. For example, for a 2-dimensional dataset, 500 queries will select 0.1\% of the domain of dimension 1, and 500 queries will select around 3.2\% of the domains of dimensions 1 and 2.

\Figure{breakdown_dimensions} shows that \TheSystem continues to outperform the baseline indexes at higher dimensions.
Note that the clustered index's relative performance also improves, since the baseline indexes spend resources on dimensions which are not frequently filtered on.
By contrast, \TheSystem learns which dimensions to prioritize; for example, on the higher-dimensional datasets, \TheSystem chooses not to include the least frequently filtered dimensions in the index. 
Yet, \TheSystem is also impacted by the curse of dimensionality (\Figure{breakdown_dimensions_ratio}), which depicts the speedup of each index compared to a full scan.
However, \TheSystem can dampen the effect of the curse through its self-optimization, degrading more slowly than other indexes.

\subsection{The Cost Model}
\label{sec:eval:num_cells}
\begin{figure}[]
    \vspace{-1em}
    \centering
    \subfloat{
        \includegraphics[width=0.5\columnwidth,clip]{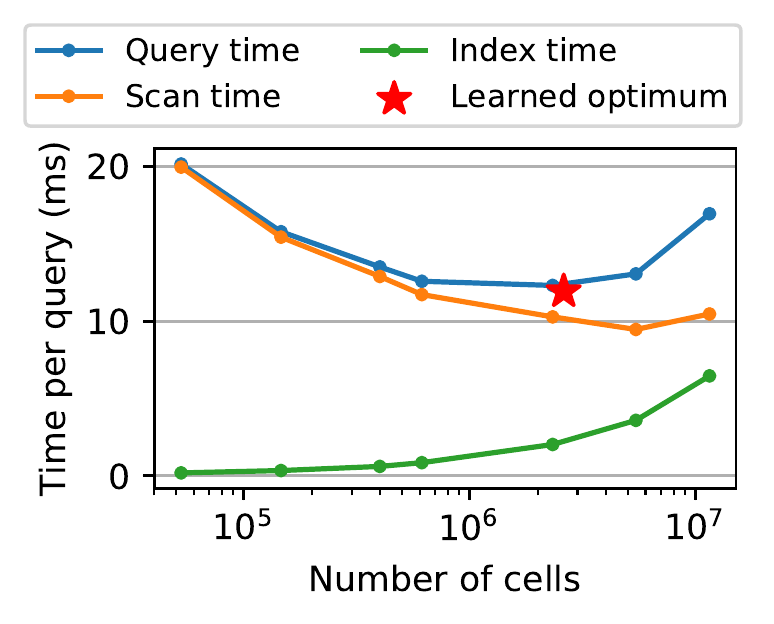}
        \label{fig:vary_num_cells_time}
        }
    ~
    \subfloat{
        \includegraphics[width=0.46\columnwidth,clip]{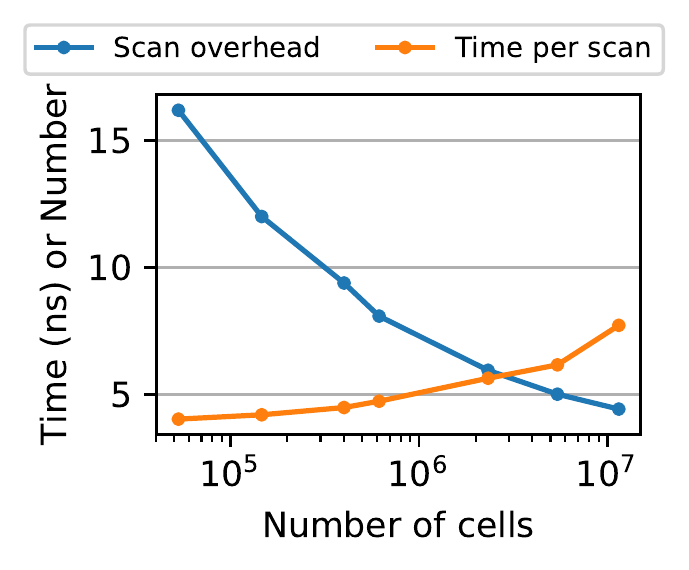}
        \label{fig:vary_num_cells_analysis1}
        }
    \vspace{-0.1in}
    \caption{TPC-H: Adding cells reduces scan overhead but incurs a higher indexing cost and worse locality.}
    \label{fig:vary_num_cells}
    \vspace{-0.1in}
\end{figure}

\NewPara{Finding the Optimum.}
Choosing an optimal layout requires balancing two competing factors: reducing the latency of locating both the relevant cells and physical index ranges within each cell (index time), and reducing the scan time by lowering scan overhead. \Figure{vary_num_cells_time} illustrates this trade-off as the size of the grid changes (we fix a layout and then scale the columns in each dimension proportionally): as the number of cells grows, scan time decreases because scan overhead decreases, but index time increases because there are more cells to process.

\TheSystem's cost model must be able to find the appropriate trade-off between scan time and index time. Indeed, \Figure{vary_num_cells_time} shows that \TheSystem finds the number of cells that minimizes the total query time (red star). Note that we show the cost surface along only a single degree of freedom for visual clarity.

\NewPara{Robustness of the model.}
As \Section{opt:calibration} describes, our cost model needs to learn the weights $\{w_p, w_l, w_r, w_s\}$, here referred to as calibration. 
This calibration should happen once per dataset and machine. 
On our server, it took around 10 minutes, most of which was spent generating training examples.

However, maybe surprisingly, the weights are quite robust to the data itself, so the cost model does not need to be retrained for every dataset.
To show this, we trained our cost model on each of our four datasets, used each model to learn layouts for all four datasets, and then ran all 16 layouts on the corresponding query workloads.
\Table{transfer_learning} shows that, no matter which dataset is used to learn the layout, the query times from the resulting layouts are similar, often with less than a 10\% difference between them.
Therefore, \TheSystem can use the same cost model regardless of changes to the dataset or query workload. 
This makes calibration a one-time cost of 10 minutes. 

\subsection{Index Creation}
\label{sec:eval:index-creation}

\begin{table}[]
\resizebox{\columnwidth}{!}{
\begin{tabular}{llllll}
\toprule
                                            &                   & \multicolumn{4}{l}{\textbf{Layout learned for}}                  \\
                                            & \textbf{}         & \textbf{sales} & \textbf{tpc-h} & \textbf{osm} & \textbf{perfmon} \\
\midrule
\multirow{4}{*}{\rotatebox[origin=c]{90}{\parbox[c]{1.1cm}{\textbf{Models trained on}}}} & \textbf{sales}    & 0.128          & 10.8 (-8\%)         & 0.975 (-7\%)       & 3.49 (+17\%)            \\
                                            & \textbf{tpch}     & 0.132 (+3\%)          & 11.7          & 0.986 (-6\%)       & 3.18 (+6\%)            \\
                                            & \textbf{osm}      & 0.134 (+5\%)         & 11.7 (+0\%)         & 1.05         & 3.14 (+5\%)            \\
                                            & \textbf{perfmon}  & 0.137 (+7\%)         & 11.6 (-1\%)        & 0.964 (-8\%)       & 2.99             \\
\bottomrule
\end{tabular}}
\centering
\caption{Query time (ms) when layouts are learned using cost models trained on different examples.}
\label{tab:transfer_learning}
\vspace{-0.2in}
\end{table}

\begin{table}[t!]
\small
\begin{tabular}{lllll}
\toprule
      & \textbf{sales} & \textbf{tpc-h}     & \textbf{osm} & \textbf{perfmon}   \\
\midrule
\TheSystem Learning & 10.3 & 33.4 & 44.5 & 33.3 \\
\TheSystem Loading & 4.12 & 29.6 & 8.03 & 22.0 \\
\textbf{\TheSystem Total} & 14.4 & 63.0 & 52.5 & 55.3 \\
\midrule
\textbf{Clustered} & 2.11 & 16.2 & 4.85 & 11.6 \\
\textbf{Z Order} & 7.82 & 86.7 & 24.9 & 72.6 \\
\textbf{UB tree} & 8.28 & 81.9 & 26.0 & 69.5 \\
\textbf{Hyperoctree} & 2.47 & 42.2 & 31.4 & 54.8 \\
\textbf{K-d tree} & 8.45 & 140 & 36.9 & 250 \\
\textbf{Grid File} & 10.6 & 121 & N/A & N/A \\
\textbf{R* tree} & 259 & N/A & 1340 & N/A \\
\bottomrule
\end{tabular}
\centering
\caption{Index Creation Time in Seconds}
\label{tab:index_creation}
\vspace{-0.3in}
\end{table}

\Table{index_creation} shows the time to create each index. We separate index creation time for \TheSystem into learning time, which is the time taken to learn the layout (\Section{opt:training}); and loading time, which is the time to build the primary index. The reported learning times use sampling of the dataset and query workload, described next. The total index creation time of \TheSystem is competitive with the creation time of the baseline indexes.

\begin{figure*}[t!]
    \centering
    \includegraphics[width=0.4\textwidth]
            {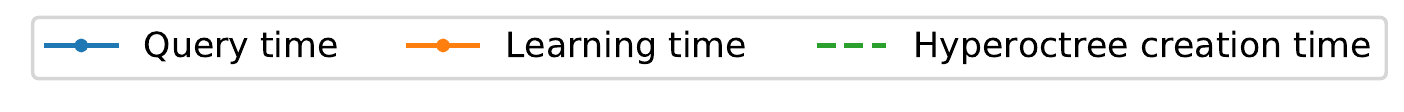}
    \\[-0.5ex] 
    \vspace{-0.2in}
    \subfloat{
        \includegraphics[width=0.24\textwidth,clip]{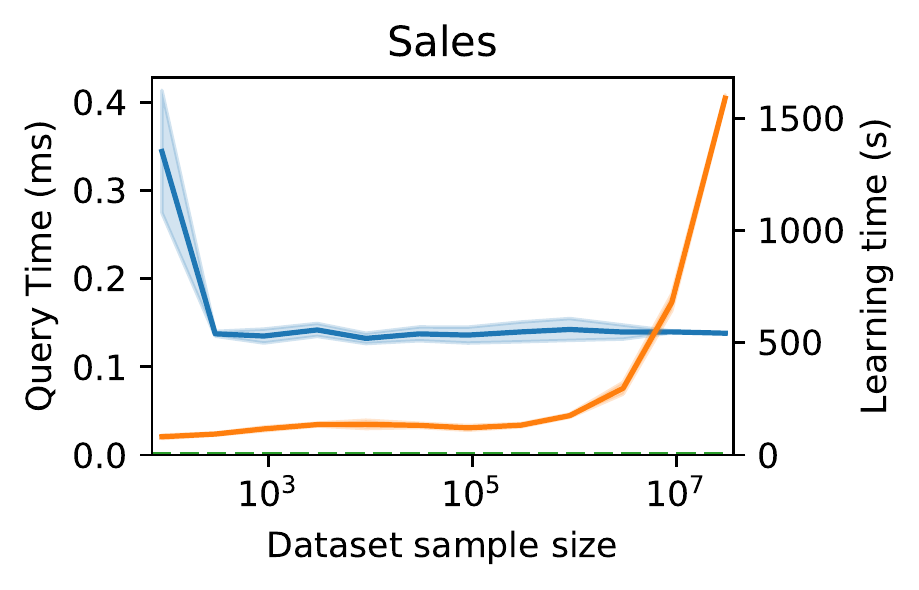}
        \label{fig:sample_dataset_ds2}
    }
    ~
    \subfloat{
        \includegraphics[width=0.24\textwidth,clip]{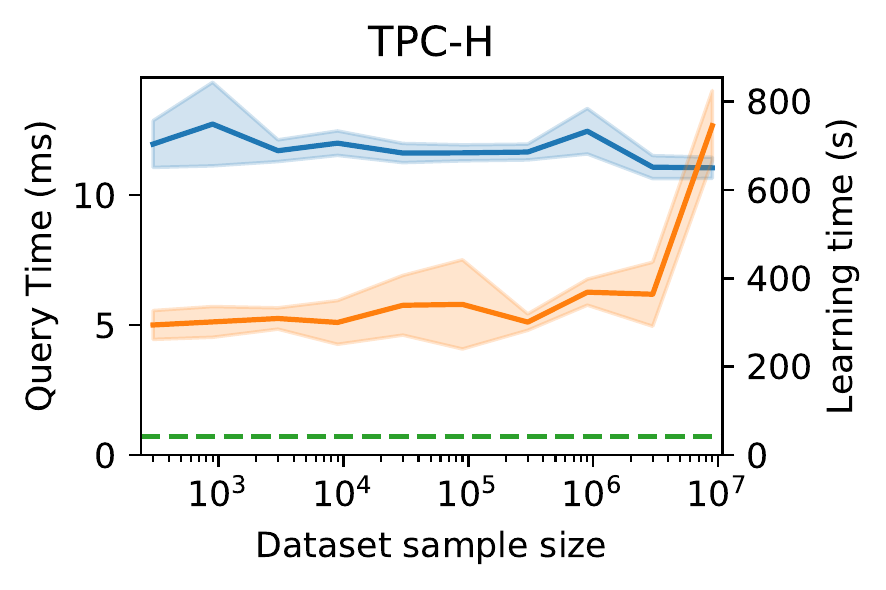}
        \label{fig:sample_dataset_tpch}
        }
    ~
    \subfloat{
        \includegraphics[width=0.24\textwidth,clip]{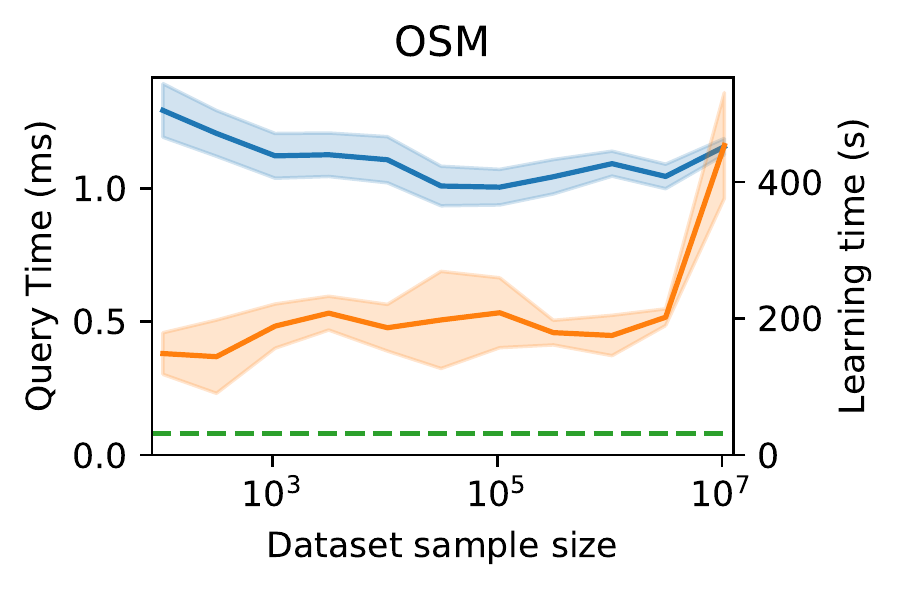}
        \label{fig:sample_dataset_osm}
        }
    ~
    \subfloat{
        \includegraphics[width=0.24\textwidth,clip]{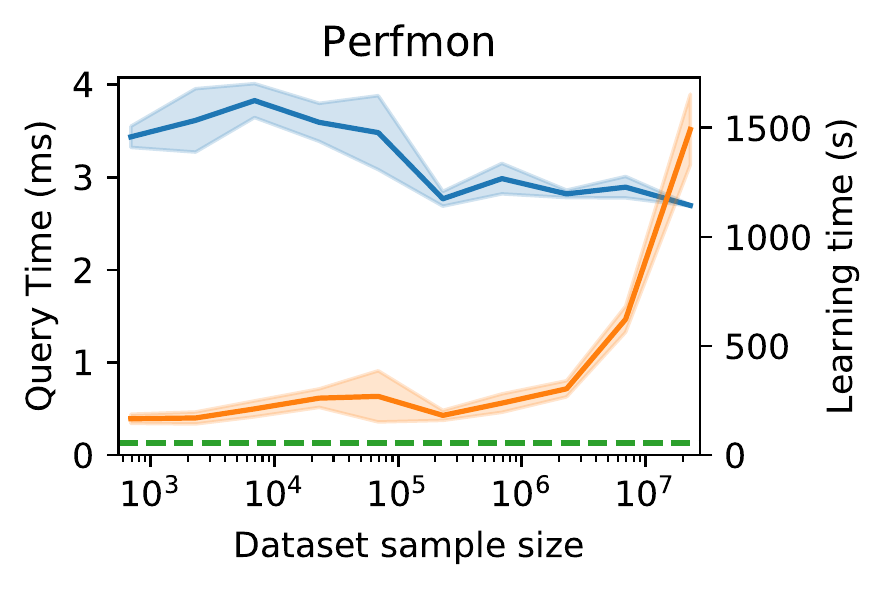}
        \label{fig:sample_dataset_perfmon}
        }
    \vspace{-0.1in}
     \caption{
        Learning time and resulting query time when sampling the dataset over several trials. One standard deviation from the mean is shaded. For comparison, we show the index creation time for the hyperoctree.
        }
        \vspace{-1em}
    \label{fig:sample_dataset}
\end{figure*}

\begin{figure*}[t!]
    \centering
    \includegraphics[width=0.4\textwidth]
            {figures/legend_sampling.pdf}
    \\[-0.5ex] 
    \vspace{-0.2in}
    \subfloat{
        \includegraphics[width=0.24\textwidth,clip]{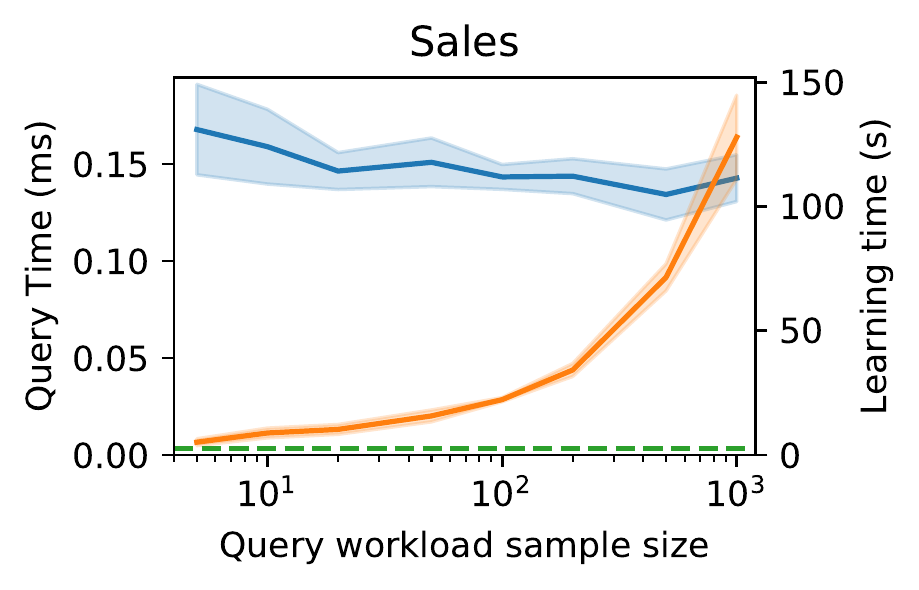}
        \label{fig:sample_workload_ds2}
    }
    ~
    \subfloat{
        \includegraphics[width=0.24\textwidth,clip]{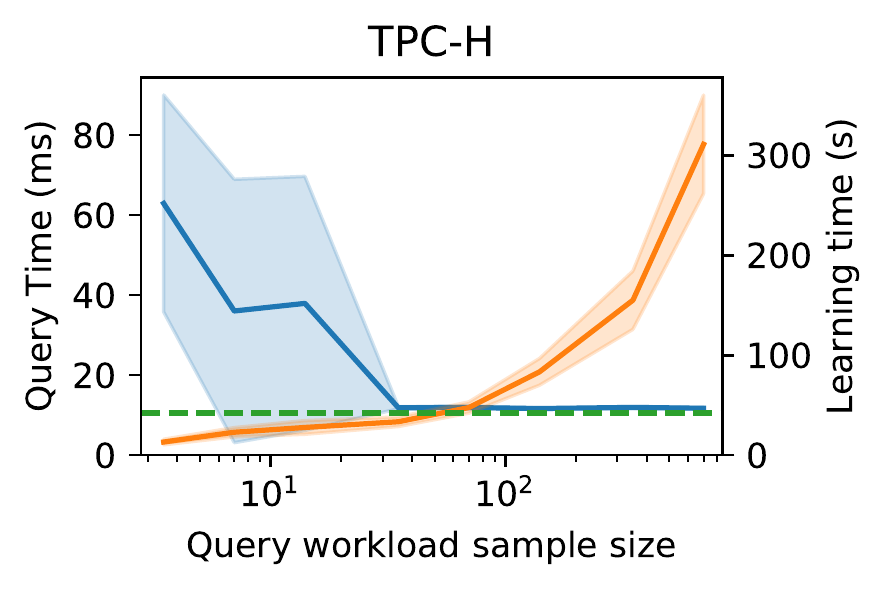}
        \label{fig:sample_workload_tpch}
        }
    ~
    \subfloat{
        \includegraphics[width=0.24\textwidth,clip]{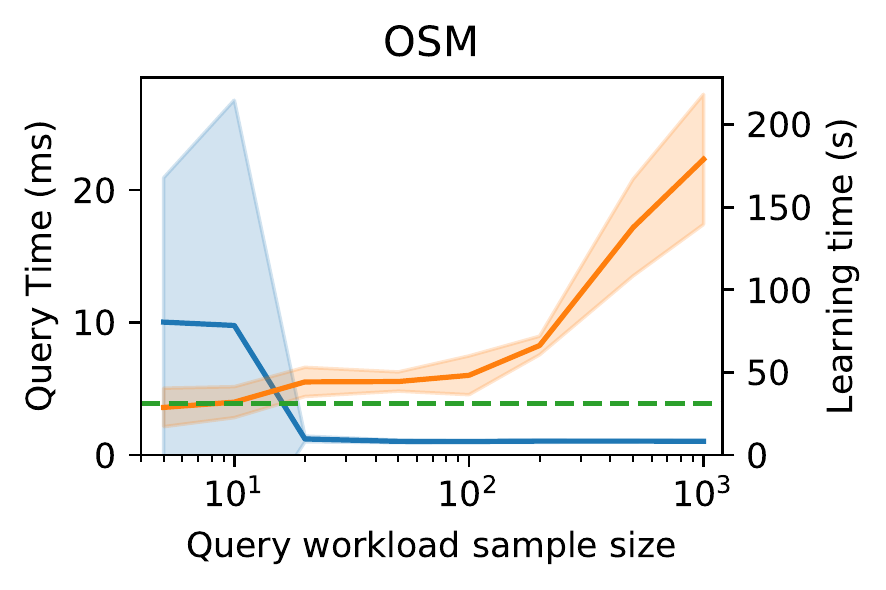}
        \label{fig:sample_workload_osm}
        }
    ~
    \subfloat{
        \includegraphics[width=0.24\textwidth,clip]{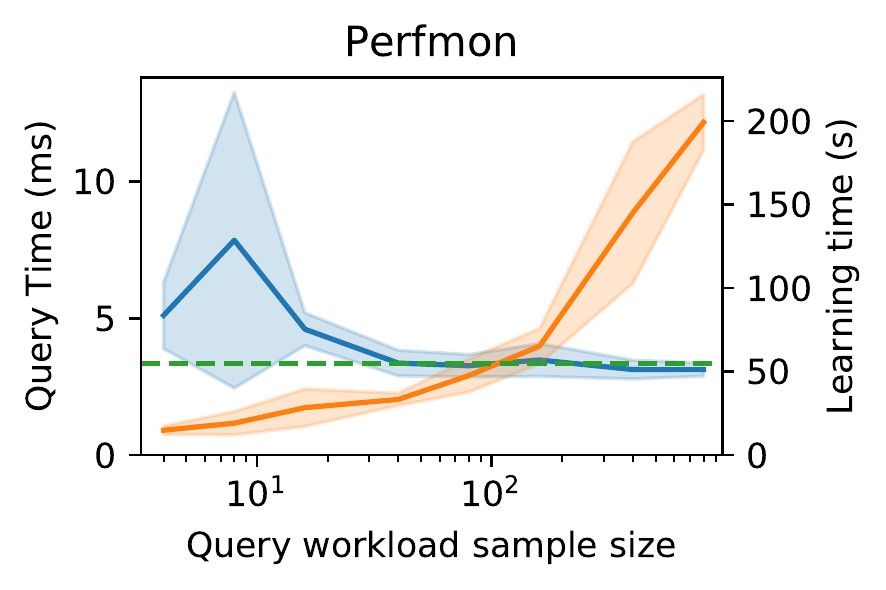}
        \label{fig:sample_workload_perfmon}
        }
    \vspace{-0.1in}
     \caption{
        Learning time and resulting query time when sampling the queries over several trials. One standard deviation from the mean is shaded. For comparison, we show the index creation time for the hyperoctree.
        }
    \label{fig:sample_workload}
\end{figure*}

\NewPara{Sampling records.}
Optimizing the layout using the entire dataset and query workload can take prohibitively long and does not scale well to larger datasets and workloads.
However, \TheSystem can reduce learning time without a significant performance loss by sampling the data.
\Figure{sample_dataset} shows that even when estimating features with a sample of only 0.01--1\%, \TheSystem maintains low query times. This is because the main purpose of the sample is to estimate the number of records scanned per query. Since our query selectivities are around 0.1\% or higher, a sample of 1\% records is sufficiently accurate.
Yet, this alone is not sufficient to match the creation time of the hyperoctree, the fastest of our multi-dimensional baselines.

\NewPara{Sampling queries.}
Sampling the query workload can reduce \TheSystem's creation time even further.
Here we conservatively use a data sample size of 100k records and vary the query sample size. 
As \Figure{sample_workload} shows, \TheSystem maintains low query times when using only 5\% of queries. This is because the query workloads contain limited number of query types. Since queries within each type have similar characteristics with respect to selectivity and which dimensions are filtered, \TheSystem only requires a few queries of each type to learn a good layout. However, the variance in performance increases as the query workload sample size decreases.
With both optimizations (data and query samples), \TheSystem achieves a learning time on par with the hyperoctree creation time without sacrificing performance.

\begin{figure}
    \centering
    \subfloat{
        \includegraphics[width=0.48\columnwidth]{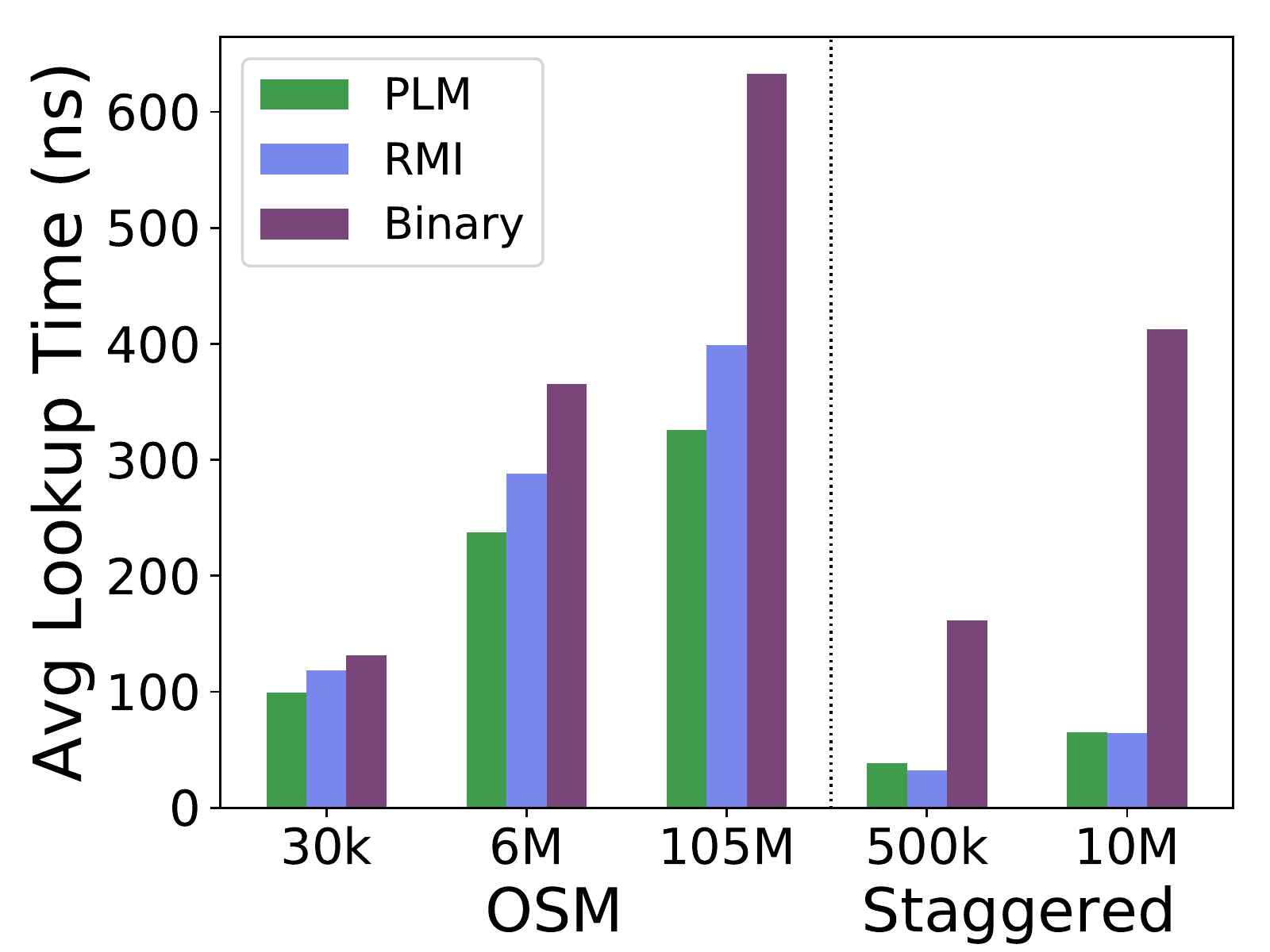}
        \label{fig:cdf-benchmark}
    }
    ~
    \subfloat{
        \includegraphics[width=0.48\columnwidth]{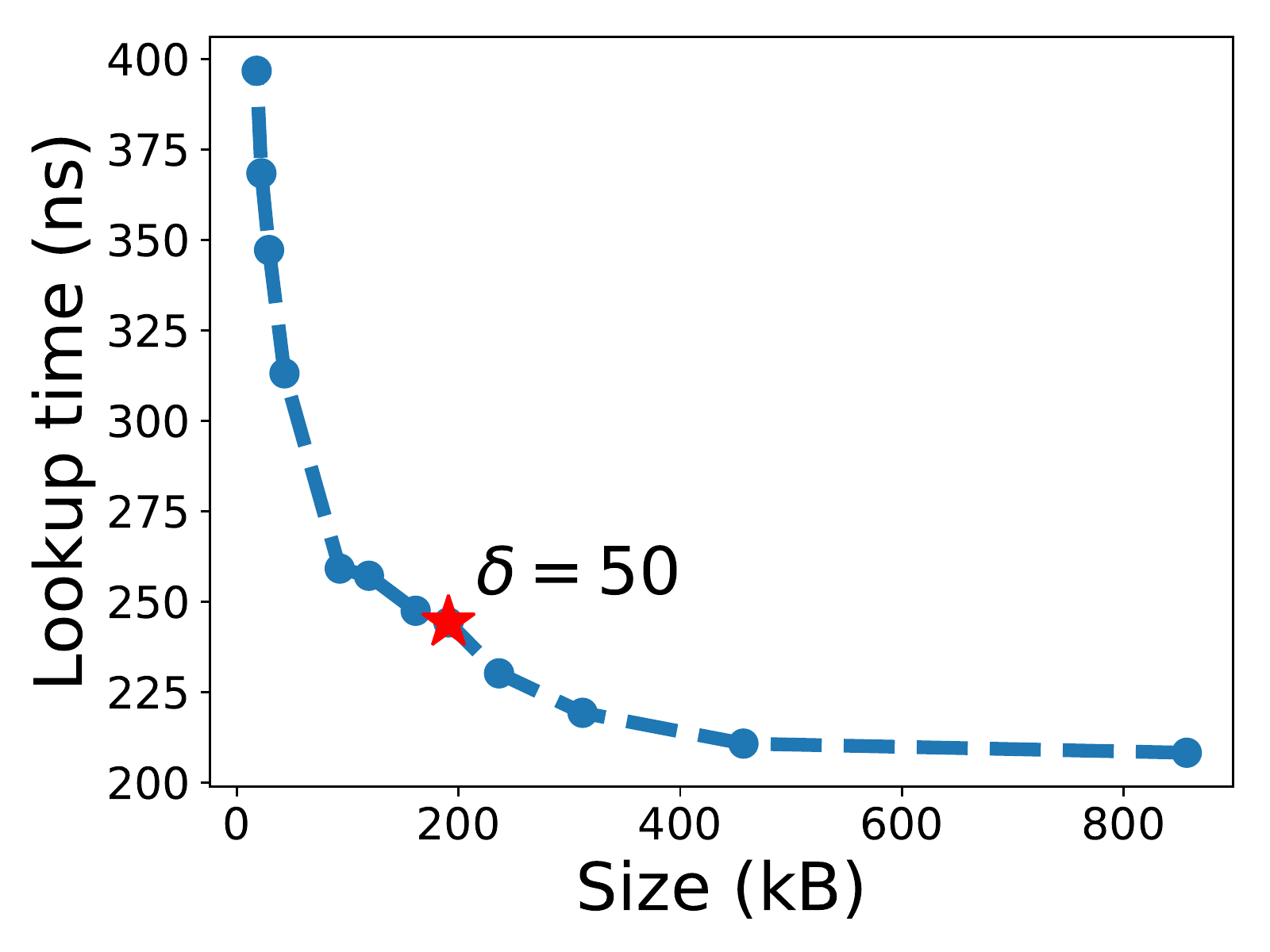}
        \label{fig:plm-tradeoff}
    }
    \vspace{-0.05in}
    \caption{(a) A comparison of three per-cell CDF models on two 1-D datasets. (b) The size-speed tradeoff for the PLM, with our configuration marked.}
    \vspace{-0.3in}
\end{figure}

\subsection{Per-cell Models}
\label{sec:eval:per-cell}
In \Section{flattening:estimate},
we discuss CDF models to accelerate the location of physical indexes along the sort dimension. Since these CDFs are evaluated twice for each visited cell (beginning and end of the range), small speedups in lookup time may be noticeable on overall query time. 
\Figure{cdf-benchmark} benchmarks the lookup time, including inference and an exponential search rectification phase, of three options we considered:
the piecewise-linear model (PLM, our approach), the learned B-tree~\cite{kraska}, and binary search. We used real data (timestamps from the OSM dataset) and synthetic staggered uniform data (data is uniform over identically sized but disjoint intervals), with query points sampled from the dataset. The PLM and RMI perform comparably, and both beat binary search by up to $4\times$ on these datasets. We use the PLM since it requires only a single tuning parameter $\delta$, which encodes the tradeoff between accuracy and size (\Figure{plm-tradeoff}). By contrast, the learned B-tree~\cite{kraska} requires extensive tuning of the number of experts per layer. The choice of $\delta$ depends on how much the administrator prioritizes speed over space: \Figure{plm-tradeoff} shows that $\delta = 50$ strikes a reasonable balance. 

\section{Future Work}
\label{sec:future}
\NewPara{Shifting workloads.} 
\TheSystem can quickly adapt to workload changes but cannot detect when the query distribution has changed sufficiently to merit a new layout. To do this, \TheSystem could periodically evaluate the cost (\Section{opt}) of the current layout on queries over a recent time window. If the cost exceeds a threshold, \TheSystem can replace the layout.
 
Additionally, \TheSystem is completely rebuilt for each new workload. However, \TheSystem could also be incrementally adjusted, e.g. by coalescing adjacent columns or splitting a column, or by incorporating aspects of incremental layout creation from database cracking~\cite{db-cracking}, to avoid rebuilding the index.

\NewPara{Insertions.} \TheSystem currently only supports read-only workloads. To support insertions, each cell could maintain gaps, similar to the fill factor of a B+Tree. It could also maintain a delta index~\cite{delta-index} in which updates are buffered and periodically merged into the data store, similar to Bigtable~\cite{bigtable}.

\NewPara{Concurrency and parallelism.} \TheSystem is currently single-threaded, but it can be extended to take advantage of concurrency and parallelism. Different cells can be refined and scanned simultaneously; within a cell, records can be scanned in parallel, allowing \TheSystem to benefit from multithreading. Additionally, since \TheSystem stores each column in the column store as a dense array, it can also take advantage of SIMD.
\vspace{-1em}
\section{Conclusion}

Despite the shift of OLAP workloads towards in-memory databases, state-of-the-art systems have failed to take advantage of multi-dimensional indexes to accelerate their queries. Many instead opt for simple 1-D clustered indexes with bulky secondary indexes that waste space. We design a new multi-dimensional index \TheSystem with two properties. First, it serves as the primary index and is used as the storage order for underlying data. Second, it is jointly optimized using both the underlying data and query workloads. Our approach outperforms existing clustered indexes by $30-400\times$. Likewise, learning the index layout from the query workload allows \TheSystem to beat optimally tuned spatial indexes, while using a fraction of the space. Our results suggest that learned primary multi-dimensional indexes offer a significant performance improvement over state-of-the-art approaches and can serve as useful building blocks in larger in-memory database systems.

\bibliographystyle{ACM-Reference-Format}
\bibliography{multi_dim_refs}

\appendix
\section{Index Implementation Details}
\label{app:implementation}
This section provides additional details for the indexes described in \Section{eval:baselines} that we implemented ourselves. When querying an index, the user provides two arguments: (1) the start and end value of the filter range in each dimension (set to negative and positive infinity if the dimension is not filtered in the query), and (2) a Visitor object which will accumulate the statistic of the aggregation. All of our experiments are performed on aggregation queries. Indexes only scan the columns for dimensions that appear in the query filter. The baseline indexes we implemented:
\begin{itemize}
    \item \textbf{Clustered Single-Dimensional Index:} We use an RMI with three layers, where all models are linear models. Models in the non-leaf layers are linear spline models to ensure that the models accessed in the following layer are monotonic; the models in the leaf layer are linear regressions. The numbers of experts in each layer are $1, \sqrt{n},$ and $n$, respectively, with $n$ tuned to minimize query time on the target workload.
    \item \textbf{Grid File:} The $d$-dimensional space is divided into \emph{blocks} by a grid (each block is one grid cell). Multiple adjacent blocks constitute a \emph{bucket}. All points in a bucket are stored contiguously and not sorted: if a record in a bucket needs to be accessed, the entire bucket must be scanned. The grid is built incrementally, starting with a single block that contains the entire space. Each point is added to its corresponding bucket; once the number of points in a bucket hits a user-defined page size, that bucket is split to form a new bucket by either (1) splitting points along an existing block boundary, if it exists in any dimension, or (2) adding a grid column (and therefore more blocks) that divides the existing bucket at its midpoint along a particular dimension. The dimension along which the block is split is cycled through in a round robin fashion. The page size is tuned to minimize query time on the target workload. When a query arrives, the grid file scans all the buckets that intersect the query rectangle.
    \item \textbf{Z-Order Index:} We use 64-bit Z-order values. When indexing $d$ dimensions, we compute the Z-order value for a point by taking the first $\lfloor 64/d \rfloor$ bits of each dimension's value and interleaving them, ordered by selectivity (e.g., the most selective dimension's LSB is the Z-order value's LSB). We order points by their Z-order value and group contiguous chunks into pages. For each page, we store the min and max value in each dimension for points in the page. Given a query, the index finds the smallest and largest Z-order value contained in the query rectangle (conceptually the bottom-left and top-right vertices of the query rectangle), uses binary search to find the physical indexes that correspond to those Z-order values, and iterates through every page that falls between those physical indexes. The points in a page are only scanned if the metadata min/max values indicate that it is possible for points in the page to match the query filter (i.e., we only scan a page if the rectangle formed by the page's min/max values intersects with the query rectangle).
    \item \textbf{UB-tree:} Z-order values are computed in the same way as the Z-Order Index. We order points by their Z-order value and group contiguous chunks into pages. For each page, we store the minimum Z-order value contained in that page. Given a query, the index finds the smallest and largest Z-order value contained in the query rectangle (conceptually the bottom-left and top-right vertices of the query rectangle), uses binary search to find the physical indexes that correspond to those Z-order values, and iterates through every physical index in this range. If we reach a Z-order value that is outside the query rectangle (the Z-order curve might enter and exit the query rectangle many times), we compute the next Z-order value that falls within the query rectangle. We then ``skip ahead'' to the page that contains this Z-order value, by comparing with each page's minimum Z-order value.
    \item \textbf{Hyperoctree:} We recursively split into $d$-dimensional hyperoctants until each page has below the page size number of points. Points within a page are stored contiguously, and pages are ordered by an in-order traversal of the tree index. Each node in the hyperoctree contains an array of points to $2^d$ child nodes, the min and max value in each dimension for points in the page, and the start and end physical index for points in the page. Given a query, the index finds all pages that intersect with the query rectangle, uses the node's metadata to identify the physical index range for each page, and scans all physical index ranges.
    \item \textbf{K-d tree:} We recursively partition space using the median value along each dimension, until the number of points in each page has below the page size number of points. The dimensions are used for partitioning in a round robin fashion, in order of decreasing selectivity. If the remaining points all have the same value in a particular dimension, that dimension is no longer used for further partitioning. Points within a page are stored contiguously, and pages are ordered by an in-order traversal of the tree index. Each node in the k-d tree contains the pointers to its two children, the dimension that is split on, the split value, and the start and end physical index for points in the page. Given a query, the index finds all pages that intersect with the query rectangle, uses the node's metadata to identify the physical index range for each page, and scans all physical index ranges.
\end{itemize}

\section{Optimization Pseudocode}
\label{app:pseudocode}
Algorithm 1 provides pseudocode for the procedure of optimizing the layout using a calibrated cost model, described in \Section{opt:training}.
\begin{algorithm*}[h]
\caption{Layout Optimization}
\begin{algorithmic}[1]
\State {\textbf{Inputs:} $d$-dimensional dataset $D$, query workload $Q=\{q_i\}$, cost model $T: (D,q,L)\rightarrow$ query time}
\State {\textbf{Output:} layout $L = ({O, C})$, where $O$ is the order of dimensions and $C = \{c_i\}_{0\le i < d-1}$ is the number columns in each grid dimension}
\Procedure{FindOptimalLayout}{$D$, $Q$, $T$}
    \State {$\widehat{D}$ = Sample($D$)}
    \State {$\widehat{Q}$ = Sample($Q$)}
    \State {/* RMIs trained on each dimension of $\widehat{D}$ are used to flatten the data and query workload samples}
    \State {\qquad by replacing each value $v$ in the $i$-th dimension of a point or query with CDF$_i(v)$ */}
    \State {$\widehat{D}, \widehat{Q}$ = Flatten($\widehat{D}, \widehat{Q}$)}
    \State {dims = /* dimensions ordered by decreasing average selectivity of $q\in\widehat{Q}$ on $\widehat{D}$ */}
    \State {best\_cost = $\infty$}
    \State {best\_L = null}
    \For{i in 0:d}
        \State{O = \{dims[0:i], dims[i+1,:], dims[i]\}}~~~~~{/* use $i$-th dimension as sort dimension */}
        \State{/* search for minimum cost $T(\widehat{D},q,(O,C))$ averaged over $q\in \widehat{Q}$, assuming fixed order O, by varying C */}
        \State{C, cost = GradientDescent(T, O, $\widehat{D}$, $\widehat{Q}$)}~~~~~{/* returns lowest found cost and the C that achieves it */}
        \State{L = (O, C)}
        \If {cost < best\_cost}
            \State {best\_cost = cost}
	        \State {best\_L = L}
        \EndIf
    \EndFor
    \State{\Return best\_L}
\EndProcedure
\end{algorithmic}
\end{algorithm*}

\end{document}